%% file: main.tex
\RequirePackage{lineno}
\documentclass[twocolumn,showpacs,superscriptaddress,amsmath,amssymb,nofootinbib]{revtex4-2}
\usepackage[x11names]{xcolor}
\usepackage{caption}
\usepackage[colorlinks, linkcolor=blue, citecolor=OliveGreen, urlcolor=magenta]{hyperref}
\usepackage{orcidlink}

\allowdisplaybreaks[4]
\usepackage{graphicx}
\usepackage{subfigure}
\usepackage{subcaption}
\usepackage{epsfig}
\usepackage{overpic}
\usepackage{dcolumn}
\usepackage{ulem}
\usepackage{bm}
\usepackage{color}
\usepackage{lineno}
\usepackage{xspace}
\usepackage{multirow}
\usepackage{epstopdf}
\usepackage{xcolor}
\usepackage{soul}
\usepackage{verbatim}
\usepackage{enumitem}
\usepackage{todonotes}
\usepackage{slashed}
\usepackage{cancel}
\usepackage{array}
\usepackage{amsmath}
\usepackage[toc,page]{appendix}

\definecolor{OliveGreen}{rgb}{0.4, 0.8, 0.1}

\usepackage{diagbox}

\newcommand{\moe}{\affiliation{Key Laboratory of Atomic and Subatomic Structure and 
Quantum Control (MOE), Guangdong-Hong Kong Joint Laboratory of Quantum Matter, Guangzhou 510006, China}}

\newcommand{\iqm}{\affiliation{State Key Laboratory of Nuclear Physics and 
Technology, Institute of Quantum Matter, South China Normal 
University, Guangzhou 510006, China}}

\newcommand{\gbrce}{\affiliation{Guangdong Basic Research Center of Excellence for 
Structure and Fundamental Interactions of Matter, Guangdong Provincial Key Laboratory of Nuclear Science, Guangzhou 510006, China}}

\newcommand{\scnt}{\affiliation{Southern Center for Nuclear-Science Theory (SCNT), Institute of Modern Physics, Chinese Academy of Sciences, Huizhou 516000, Guangdong Province, China}}

\newcommand{\OU}{\affiliation{Research Center for Nuclear Physics (RCNP), Osaka University, Ibaraki 567-0047, Japan}}

\begin{document}
\include{def-com}
\title{\boldmath Why is the $Z_c(3900)$ absent in the $h_c\pi$ final state? }

\author {\mbox{Quanxing Ye}\orcidlink{0009-0006-8322-3163}}
\iqm
\moe
\gbrce

\author {\mbox{Ying Zhang\orcidlink{0009-0008-9322-1625}}}
\email{Co-first author}
\iqm
\moe
\gbrce

\author {\mbox{Peng-Yu Niu\orcidlink{0000-0001-8455-9570}}}
\email{niupy@m.scnu.edu.cn}
\iqm
\gbrce

\author {\mbox{Qian Wang\orcidlink{0000-0002-2447-111X}}}
\email{qianwang@m.scnu.edu.cn, corresponding author}
\iqm
\gbrce
\scnt
\OU

\date{\today}
\input{Abstract}

\maketitle

\input{Introduction}
\input{Formalism}

\input{Results}
\input{Summary}

\section*{Acknowledgments}
We are grateful to Meng-Lin Du and Feng-Kun Guo for the helpful discussion. 
This work is partly supported by the National Natural Science Foundation of China with Grants No.~12375073, No.~12547105, and No.~12505110.

\nocite{*}
\bibliography{ref.bib}  

\onecolumngrid
\newpage
\appendix


\input{supp}

\end{document}

%% file: Abstract.tex
\begin{abstract}
In this work, we perform a comprehensive phenomenological analysis of the exotic hadronic states $Z_c(3900)$, $Z_c(4020)$, $Z_b(10610)$ and $Z_b(10650)$ within the framework of Heavy Quark Spin Symmetry (HQSS) and its violation. By constructing S-wave contact interactions between elastic ($D\bar{D}^*/D^*\bar{D}^*$ or $B\bar{B}^*/B^*\bar{B}^*$) and inelastic ($J/\psi\pi, h_c\pi$ or $\Upsilon\pi$, $h_b\pi$) channels, we solve the Lippmann-Schwinger equation to obtain physical production amplitudes and perform a global fit to experimental invariant-mass spectra. Our results demonstrate a striking difference between the charm and bottom sectors: HQSS violation is negligible in the bottom system, leading to comparable peak structures for both $Z_b$ states in all hidden-bottom decay channels. In contrast, significant HQSS breaking is required to describe the $Z_c$ system, where the violation is predominantly concentrated in the elastic interactions. This explains the observed selectivity: $Z_c(3900)$ appears prominently only in $J/\psi\pi$, while $Z_c(4020)$ appears only in $h_c\pi$. Pole analysis confirms the molecular nature of the states, with the $Z_c(4020)$ likely arising  from a threshold cusp effect. The model's robustness is verified against variations of the form factor and cutoff, showing stable results.
 
\end{abstract}

%% file: Introduction.tex
\section{INTRODUCTION}
The past decades have witnessed the discovery of numerous exotic states, particularly charmonium-like and bottomonium-like states. Typical examples include the charmonium-like $Z_c(3900)$~\cite{Xiao:2013iha, BESIII:2013ris, BESIII:2013qmu, BESIII:2015pqw, BESIII:2017bua, Belle:2013yex} and $Z_c(4020)$ ~\cite{ BESIII:2013ouc, BESIII:2013mhi, BESIII:2015tix} states, as well as the bottomonium-like $Z_b(16610)$ and $Z_b(10650)$  states~\cite{Belle:2011aa,Belle:2014vzn,Belle:2015upu}, which carry exotic flavor quantum numbers and therefore cannot be accommodated in the conventional quarkonium picture due to their charged property but with a pair of heavy quarks. 

The two narrow bottomonium-like structures, e.g. $Z_b(10610)$ and $Z_b(10650)$, were reported by Belle Collaboration in both  
$\Upsilon(nS)\pi^{\pm}$ ($n=1,2,3$) and $h_b(mP)\pi^{\pm}$ ($m=1,2$) invariant-mass spectra~\cite{Belle:2011aa}. Both structures were subsequently confirmed in the $B^{(*)}\bar B^{(*)}$ channels~\cite{Belle:2015upu}. Their masses lie extremely close to the $B\bar{B}^*$ and $B^*\bar{B}^*$ thresholds, respectively, suggesting a natural interpretation as $B^{*}\bar{B}^{(*)}$ molecular states. Moreover, these two peaks exhibit comparable heights in hidden bottom channels~\cite{Belle:2011aa, Belle:2014vzn}. That is in qualitative agreement with the expetations of heavy quark spin symmetry~\cite{Bondar:2011ev,Wang:2018jlv,Baru:2019xnh}, i.e. the two $Z_b$ states couple equally to a given hidden bottom channel.

If Heavy Quark Flavor symmetry (HQFS) and HQSS were perfectly preserved, one would similarly expect two nearly similar structures in the charmonium-like channels~\cite{Guo:2013sya}, especially in the $J/\psi\pi^{\pm}$, $h_c\pi^{\pm}$ channels. 
 However, current measurements reveal only one prominent peak around the $D^{*}\bar{D}$ threshold in the $J/\psi\pi^{\pm}$ invariant mass distribution~\cite{Xiao:2013iha, BESIII:2013ris, BESIII:2013qmu} and one around the $D^*\bar{D}^*$ threshold in the $h_c\pi^{\pm}$~\cite{BESIII:2013ouc} invariant mass distribution, corresponding to $Z_c(3900)$ and $Z_c(4020)$, respectively. Although, those two $Z_c$ states are analogies of the two $Z_b$ states, by replacing $b$ ($\bar{b}$) by $c$ ($\bar{c}$) in both compact tetraquark picture~\cite{Ali:2013xba, Maiani:2013nmn, Voloshin:2013dpa, Ali:2014dva, Patel:2014vua, Deng:2014gqa, Deng:2015lca, Voloshin:2018pqn, Wang:2022clw, Wua:2023ntn, Junnarkar:2024kwd, Ghasempour:2025luy, Kang:2025xqm} and hadronic molecular picture~\cite{Chen:2016qju, Lebed:2016hpi, Esposito:2016noz, Chen:2016spr, Guo:2017jvc, Olsen:2017bmm, Brambilla:2019esw, Yang:2020nrt, Baru:2020ywb, Wu:2022hck, Chen:2022asf, Meng:2022ozq, Yildirim:2023znd, Wu:2023rrp, Liu:2024uxn}, 
 it is still an open question why the lower (higher) $Z_c$ state only exhibits significant structure in the $J/\psi\pi$ ($h_c\pi$) channel. 
 This qualitative difference between the $Z_b$ and $Z_c$ sectors indicates a potentially nontrivial pattern of HQSS breaking, which is expected to be more significant in the charmonium sector due to the smaller charm quark mass. This is the key to lead to different behaviors of the $Z_c(3900)$ and $Z_c(4020)$ in the $J/\psi\pi$ and $h_c\pi$ channels. 

In principle, HQSS imposes strong constraints on systems containing a single heavy quark. For systems involving two heavy hadrons, however, these symmetries are violated, with HQSS receiving corrections at next-to-leading order in the heavy-quark expansion. Hence, a quantitative investigation of HQSS breaking effects in both the $Z_c$ and $Z_b$ systems is crucial for clarifying the underlying structure of the observed states and assessing the molecular interpretation. It should be emphasized that structures appearing near open-flavor thresholds may originate not only from genuine bound, virtual or resonant states \cite{Albaladejo:2015lob, Gong:2016hlt, He:2017lhy, Du:2022jjv}, but also from kinematic effects \cite{Swanson:2014tra, Guo:2019twa, Dong:2020hxe}. These effects make the line shapes highly sensitive to coupled-channel dynamics and analyticity constraints. Consequently, identifying the nature of these states necessitates a dynamical treatment via the Lippmann–Schwinger equation (LSE) and a systematic examination of regulator and form-factor dependence.

This work is organized as follows. In Sec.~\ref{Formalism}, we construct the contact potentials for the $Z_c$ and $Z_b$ systems based on HQSS with its breaking effects included, and then insert them into the LSE to obtain the production amplitudes. Then we adopt the differential decay rate for a three-body decay to obtain our fitting function.
In Sec.~\ref{Results}, we present the fit results in the $Z_c$ and $Z_b$ systems, and extract the poles in the energy complex plane and the corresponding effective couplings. The sensitivity of our model to the choice of form factors is also tested. Sec.~\ref{Summary} is a brief summary.
Some calculation details are presented in Appendixes.

%% file: Formalism.tex
\section{FORMALISM}\label{Formalism}

In this work, we work on the leading order short-ranged contact potentials constructed from the heavy-light decomposition proposed by Voloshin~\cite{Bondar:2011ev} in the heavy-quark limit, which is similar to the so-called pionless nuclear effective field theory~\cite{Bedaque:2002mn,Epelbaum:2008ga}. In the heavy-light decomposition, the HQSS breaking contact potentials can also be constructed. In this section, we use  
 the $Z_c$ system as an example to construct the leading-order contact potentials with HQSS breaking effect explicitly included. The potentials of the $Z_b$ system are analogous.

\subsection{The contact potentials of $Z_c$ and $Z_b$ systems}\label{p}
Within the molecular framework, the $Z_c(3900)$ and $Z_c(4020)$ states can be expressed as
\begin{align}\nonumber
    |Z_c(3900)\rangle & = |D\bar{D}^*-c.c.\rangle_{1^{+-}},~~|Z_c(4020)\rangle  = |D^*\bar{D}^*\rangle_{1^{+-}},
\end{align}
based on their approaching to the $D\bar{D}^*$ and $D^*\bar{D}^*$ thresholds, respectively. 
The relative negative sign in $|D\bar{D}^*-c.c.\rangle_{1^{+-}}$ is due to the negative charged parity of the $Z_c(3900)$, with $D^{(*)}\rightarrow\bar{D}^{(*)}$ under charge transformation. Throughout this work, we adopt the convention 
\begin{align}
    |\mathcal{C}=\pm\rangle = \frac{1}{\sqrt{2}}[AB \pm (-1)^{{j-j_A-j_B}}\bar{B}\bar{A}], 
\end{align}
where $j_A$, $j_B$, and $j$ denote the spin of meson A, the spin of meson B, and their total spin, respectively~\cite{Guo:2017jvc}. 

Starting from the particle basis, one can perform a heavy–light decomposition 
to construct the contact potentials for the  $Z_c$ system. Specifically, the particle basis $|\left[s_{l_{1}} s_{Q_{1}}\right]_{j_{1}}\left[s_{l_{2}} s_{Q_{2}}\right]_{j_{2}}\rangle_{J}$ can be reorganized into the heavy–light basis $|\left[s_{Q_{1}} s_{Q_{2}}\right]_{s_{Q}}\left[s_{l_{1}} s_{l_{2}}\right]_{s_{l}}\rangle_{J}$  
, which is denoted as $|s_Q \otimes s_l\rangle_J$ for brevity.
Here, $s_{l_i}$ is the light degree of freedom of the $i$-th heavy-light meson, i.e., the light quark spin plus the relative orbital angular momentum $l$.  $s_{Q_i}$ and $j_i$ are  
the heavy-quark spin and the total spin of the $i$-th meson, respectively. Under these conventions, the decomposition relation reads as
\begin{align}
|[s_{Q_1}s_{l_1}]_{j_1}[s_{Q_2}s_{l_2}]_{j_2}\rangle_J
= & \sum_{s_l,s_Q}\hat{s_Q}\,\hat{s_l}\,\hat{j_1}\,\hat{j_2} 
\left\{
\begin{array}{ccc}
s_{Q_1} & s_{Q_2} & s_Q \\
s_{l_1} & s_{l_2} & s_l \\
j_1 & j_2 & J
\end{array}
\right\}\notag\\ 
& \times |s_Q \otimes s_l \rangle_J, 
\end{align}
with $\hat{j}=\sqrt{2j+1}$. 
Therefore, the wave functions of the  $Z_c(3900)$ and $Z_c(4020)$ systems in the heavy-light basis are~\cite{Bondar:2011ev} 
\begin{align}
|Z_c(3900)\rangle & = |D\bar{D}^*-c.c.\rangle_{1^{+-}}= - \frac{1}{\sqrt{2}}|1\otimes 0\rangle-\frac{1}{\sqrt{2}}|0\otimes1\rangle,\notag\\
|Z_c(4020)\rangle & = |D^*\bar{D}^*\rangle_{1^{+-}}=\frac{1}{\sqrt{2}}|1\otimes 0\rangle-\frac{1}{\sqrt{2}}|0\otimes1\rangle.
\label{hld}
\end{align}

In the heavy quark limit, the heavy and light degrees of freedoms are conserved, respectively \cite{Neubert:1993mb},
with the dynamics only dependent on the light degree of freedom. 
In this case, one can define the low-energy parameters 
\begin{eqnarray}
    \mathcal{C}_0\equiv \langle  1\otimes 0|\hat{\mathcal{H}}|1\otimes 0\rangle,\notag\\
     \mathcal{C}_1\equiv \langle  0\otimes 1|\hat{\mathcal{H}}|0\otimes 1\rangle,
     \label{HQSC}
\end{eqnarray}
where $\hat{\mathcal{H}}$ denotes the leading-order Hamiltonian density that respects HQSS. Similarly, the low-energy coefficients for HQSS breaking can be defined as 
\begin{eqnarray}
    \mathcal{B}\equiv \langle  1\otimes 0|\hat{\mathcal{H}'}|0\otimes 1\rangle= \langle  0\otimes 1|\hat{\mathcal{H}'}|1\otimes 0\rangle,
    \label{eb}
\end{eqnarray}
where $\hat{\mathcal{H}'}$ represents the leading-order Hamiltonian density for HQSS breaking effect.
With the above definitions, 
the potentials for the $Z_c$ system reads as 
\begin{eqnarray}
V_{\text{e}}^\text{c}=\left(
\begin{array}{cc}
    \frac{1}{2}(\mathcal{C}_0+\mathcal{C}_1) + \mathcal{B} & \frac{1}{2}(\mathcal{C}_1-\mathcal{C}_0)  \\
    \frac{1}{2}(\mathcal{C}_1-\mathcal{C}_0) &   \frac{1}{2}(\mathcal{C}_0+\mathcal{C}_1) - \mathcal{B}
\end{array}
\right ).
\label{vct}
\end{eqnarray}
with the $D\bar{D}^*+c.c.$ and $D^*\bar{D}^*$ the corresponding channels. 

For the $Z_c$ case, we also consider two inelastic channels, i.e., $J/\psi\pi$ and $h_c\pi$. Since the two-body system $J/\psi\pi$ and $h_c(1P)\pi$ can be expressed as $|1\otimes 0\rangle$ and $|0\otimes 1\rangle$, respectively, in the heavy-light basis, they can couple to the $|1\otimes 0\rangle$ and $|0\otimes 1\rangle$ bases 
\begin{eqnarray}
    g^\psi_{1S}&\equiv& \langle \psi(1S)\pi|\hat{\mathcal{H}}_{\text{in}}|1\otimes 0\rangle,\notag\\
    g^{h_c}_{1P}&\equiv& \langle h_c(1P)\pi|\hat{\mathcal{H}}_{\text{in}}|0\otimes 1\rangle,
\end{eqnarray}
respectively, via HQSS-preserving interaction $\hat{\mathcal{H}}_\text{in}$.
One can also define HQSS breaking couplings 
between the inelastic channels and the elastic channels
\begin{eqnarray}
    v^\psi_{1S}&\equiv& \langle \psi(1S)\pi|\hat{\mathcal{H}'}_{\text{in}}|0\otimes 1\rangle,\notag\\
    v^{h_c}_{1P}&\equiv& \langle h_c(1P)\pi|\hat{\mathcal{H}'}_{\text{in}}|1\otimes 0\rangle,
    \label{ieb}
\end{eqnarray}
via HQSS breaking interactions 
$\hat{\mathcal{H}}_{\text{in}}^\prime$. In this case, the contact potential matrix between the  elastic channels and the  inelastic channels can be written as
\begin{align}
    V_{\text{in}}^{\text{c}}=\left(\begin{array}{cc}
        -\frac{1}{\sqrt{2}}g_{1S}^{\psi}-\frac{1}{\sqrt{2}}v_{1S}^{\psi} & \frac{1}{\sqrt{2}}g_{1S}^{\psi}-\frac{1}{\sqrt{2}}v_{1S}^{\psi}  \\
        -\frac{1}{\sqrt{2}}g_{1P}^{h_c} -\frac{1}{\sqrt{2}}v_{1P}^{h_c} & -\frac{1}{\sqrt{2}}g_{1P}^{h_c} +\frac{1}{\sqrt{2}}v_{1P}^{h_c}
    \end{array}\right).
\end{align}

The complete potential in the four-channel system, i.e., $D\bar{D}^*+c.c., D^*\bar{D}^*, \psi(1S)\pi, h_c(1P)\pi$ in order, is given by
\begin{align}
    V^{\text{c}} = \left( \begin{array}{cc}
        V_{\text{e}}^\text{c} & V_{\text{in}}^{\text{c}~\text{T}} \\
        V_{\text{in}}^{\text{c}}  & \bm{0}_{2\times 2}
    \end{array}\right),
    \label{vcharm}
\end{align}
where $V_{\text{in}}^{\text{c}~\text{T}}$ is the transpose matrix of $ V_{\text{in}}^{\text{c}}$. Here, the transition between the inelastic channels is negligible, as the direct interactions between heavy quarkonia and pions are expected to be suppressed. Since charmonium and bottomonium do not contain light quarks, their interaction with pions are suppressed due to the OZI suppression, which is consistent with the conclusion from both the EFT~\cite{Liu:2012dv} and lattice calculations ~\cite{Detmold:2012pi}. Thus, it is reasonable to set the contact potential between inelastic channels to zero as those in Res.~\cite{Hanhart:2015cua, Guo:2016bjq, Zhang:2025fcv}.

The contact potentials of the $Z_b$ system can also be obtained analogously. For this system, we consider $e^+e^-\to B\bar{B}^*\pi$, $B^*\bar{B}^*\pi$ two elastic channels, and $e^+e^-\to \Upsilon(1S)\pi^+\pi^-$, $\Upsilon(2S)\pi^+\pi^-$, $h_b(1P)\pi^+\pi^-$ and $h_b(2P)\pi^+\pi^-$ four inelastic channels. The $\Upsilon(3S)\pi^+\pi^-$ channel is not considered, as the $Z_b^+$ in the $\Upsilon(3S)\pi^+$ invariant mass distribution and the $Z_b^-$ in the $\Upsilon(3S)\pi^-$ invariant mass distribution may overlap with each other, which needs to dealt with three-body final state interaction rigorously.
Accordingly, the expression of the contact potential  $V_{\text{e}}^\text{c}$ defined in Eq.~(\ref{vct}) can be directly applied to the $Z_b$ system, denoted as $V_{\text{e}}^{\text{b}}$. Analogously, the HQSS-conserving coupling constants between  the inelastic channels and the elastic channels are defined as 
\begin{align}
    g^\Upsilon_{1S}&\equiv \langle \Upsilon(1S)\pi|\hat{\mathcal{H}}_{\text{in}}|1\otimes 0\rangle,~
    g^{\Upsilon}_{2S}\equiv \langle \Upsilon(2S)\pi|\hat{\mathcal{H}}_{\text{in}}|1\otimes 0\rangle,\notag\\
    g^{h_b}_{1P}&\equiv \langle h_b(1P)\pi|\hat{\mathcal{H}}_{\text{in}}|0\otimes 1\rangle,~
    g^{h_b}_{2P} \equiv \langle h_b(2P)\pi|\hat{\mathcal{H}}_{\text{in}}|0\otimes 1\rangle.
\end{align}
The HQSS breaking coupling constants are similarly defined as 
\begin{align}
    v^\Upsilon_{1S}&\equiv \langle \Upsilon(1S)\pi|\hat{\mathcal{H}'}_{\text{in}}|0\otimes 1\rangle,~
    v^{\Upsilon}_{2S}\equiv \langle \Upsilon(2S)\pi|\hat{\mathcal{H}'}_{\text{in}}|0\otimes 1\rangle,\notag\\
    v^{h_b}_{1P}&\equiv \langle h_b(1P)\pi|\hat{\mathcal{H}'}_{\text{in}}|1\otimes 0\rangle,~
    v^{h_b}_{2P} \equiv \langle h_b(2P)\pi|\hat{\mathcal{H}'}_{\text{in}}|1\otimes 0\rangle.
\end{align}
Hence, the potential matrix between the inelastic channels and the elastic channels in the $Z_b$ system can be expressed by
\begin{align}
    V_{\text{in}}^{\text{b}} = \left(\begin{array}{cc}
         -\frac{1}{\sqrt{2}}g_{1S}^{\Upsilon}-\frac{1}{\sqrt{2}}v_{1S}^{\Upsilon} & \frac{1}{\sqrt{2}}g_{1S}^{\Upsilon}-\frac{1}{\sqrt{2}}v_{1S}^{\Upsilon} \\
         -\frac{1}{\sqrt{2}}g_{2S}^{\Upsilon}-\frac{1}{\sqrt{2}}v_{2S}^{\Upsilon}&  \frac{1}{\sqrt{2}}g_{2S}^{\Upsilon}-\frac{1}{\sqrt{2}}v_{2S}^{\Upsilon}\\
         -\frac{1}{\sqrt{2}}g_{1P}^{h_b}-\frac{1}{\sqrt{2}}v_{1P}^{h_b}  &  -\frac{1}{\sqrt{2}}g_{1P}^{h_b}+\frac{1}{\sqrt{2}}v_{1P}^{h_b} \\
         -\frac{1}{\sqrt{2}}g_{2P}^{h_b}-\frac{1}{\sqrt{2}}v_{2P}^{h_b}  & -\frac{1}{\sqrt{2}}g_{1P}^{h_b}+\frac{1}{\sqrt{2}}v_{1P}^{h_b} 
    \end{array}\right).
\end{align}
Finally, the complete potential of the six-channel system, which involves two elastic and four inelastic channels, takes the form
\begin{align}
    V^{\text{b}} = \left( \begin{array}{cc}
        V_{\text{e}}^\text{b} & V_{\text{in}}^{\text{b}~\text{T}} \\
        V_{\text{in}}^{\text{b}}  & \bm{0}_{4\times 4}
    \end{array}\right),
    \label{potential}
\end{align}
where $V_{\text{in}}^{\text{b}~\text{T}}$ denotes the transpose matrix of $ V_{\text{in}}^{\text{b}}$.

\subsection{The Lippmann-Schwinger equation}
The full $T$-matrix is obtained by solving LSE 
\begin{align}
    T(E)=V+VG(E)T(E),
    \label{LSE}
\end{align}
where $V$ and $G(E)$ denote the full potential matrix and the two-point loop function matrix, respectively, in the $Z_c$ or $Z_b$ systems. The explicit form of the loop function matrix is given by
\begin{align}
    G(E)=\left(\begin{array}{cc}
        \text{diag}[G_{\text{e}}(E)]_{n\times n} & \bm{0}_{n\times m} \\
         \bm{0}_{m\times n}& \text{diag}[G_{\text{in}}(E)]_{m\times m} 
    \end{array}\right),
    \label{G}
\end{align}
where $n$ and $m$ are the number of the elastic and inelastic channels, respectively. 
Here, $(n,m)=(2,2)$ for the $Z_c$ case and $(n,m)=(2,4)$ for the $Z_b$ one. The two-body propagator for the elastic channel reads
\begin{align}
    G_{\text{e}}(E)& = \int \frac{d^{3} q}{(2 \pi)^{3}} \frac{ {f_{\Lambda}^2\left(\vec{q}\right)}}{E-m_{1}-m_{2}-\vec{q}^{2} /(2 \mu)+i\varepsilon^+}\notag\\
    & = -\frac{\mu\Lambda}{(2\pi)^{3/2}}+\frac{\mu k}{2\pi}e^{-2k^2/\Lambda^2}\left[\mathrm{erfi}\left(\frac{\sqrt{2}k}{\Lambda}\right)-i\right],
    \label{G_fun}
\end{align}
where $k=\sqrt{2\mu(E-m_1-m_2)}$ with $\mu=m_1m_2/(m_1+m_2)$ is the three-momentum of the open charm (bottom) meson in the center-of-mass (c.m.) frame of the corresponding meson pair. In Eq.~(\ref{G_fun}), the Gaussian form factor $f_{\Lambda}(\vec{q})=\text{exp}(-\Vec{q}^2/\Lambda^2)$ is introduced with cutoff $\Lambda$, and $\text{erfi}(x)=\frac{2}{\sqrt{\pi}}\int_{0}^{z}e^{t^2}dt$ is the imaginary error function. $G_{\mathrm{in}}$ is the two-body propagator of the inelastic channels. Its form will be given later.

Substituting Eq.~(\ref{vcharm}) (or Eq.~(\ref{potential})) and Eq.~(\ref{G}) into Eq.~(\ref{LSE}), one can obtain 
\begin{widetext}
\begin{align}
\left(\begin{array}{cc}
\left[T_{\text{ee}} \right]_{n \times n}& \left[T_{\text{ei}} \right]_{n \times m}\\
\left[T_{\text{ie}} \right]_{m \times n}& \left[T_{\text{ii}}\right]_{m\times m}
\end{array}\right) & = \left(\begin{array}{cc}
\left[V_{\text{e}}+V_{\text{e}} G_{\text{e}} T_{\text{ee}}+V_{\text{in}}^\text{T} G_{\text{in}} T_{\text{ie}}\right]_{n\times n} & \left[V_{\text{in}}^\text{T}+V_{\text{e}} G_{\text{e}} T_{\text{ei}}+V_{\text{in}}^\text{T} G_{\text{in}} T_{\text{ii}}\right]_{n\times m} \\
\left[V_{\text{in}}+V_{\text{in}} G_{\text{e}} T_{\text{ee}}\right]_{m\times n} & \left[V_{\text{in}} G_{\text{e}} T_{\text{ei}}\right]_{m\times m}
\end{array}\right),
\label{T}
\end{align}
\end{widetext}
By plugging the $T_{\text{ie}}$ into $T_{\text{ee}}$, one obtains
\begin{align}
T_{\text{ee}}(E)=\hat{V}_{\text{e}}^{\text{eff}}(E)+\hat{V}_{\text{e}}^{\text{eff}}(E)G_{\text{e}}(E) T_{\text{ee}},
\label{t_matrix}
\end{align}
where the effective potential is defined as $\hat{V}_{\text{e}}^{\text{eff}}\equiv V_{\text{e}}+V_{\text{in}}^{\text{T}}G_{\text{in}}V_{\text{in}}$.
From the definition of the effective potential $\hat{V}_{\text{e}}^{\text{eff}}$, it is evident that $G_{\text{in}}$ does not appear explicitly, and could be viewed as part of polynomial combinations within $\hat{V}_{\text{e}}^{\text{eff}}$. Consequently, the real part of $G_{\text{in}}$ can always be absorbed into the definition of  $V_{\text{e}}$, and only its imaginary part needs to be explicitly retained. Since the energy range under consideration in this work is relatively narrow, the coefficients in the imaginary part can be treated as energy independent. Therefore, the two-body propagator for the inelastic channels $G_{\text{in}}$ can be defined as~\cite{Hanhart:2015cua,Wang:2018jlv} 
\begin{align}
    G_{\text{in}}(S\text{-wave}) =ip,\quad G_{\text{in}}(P\text{-wave}) = ip^3,
\end{align}
where $p=\sqrt{2\mu(E-E_{\text{thr}})}$ with $E_\text{thr}$ and $\mu$ the threshold and reduced mass of the corresponding inelastic channel. 

To ensure the unitarity of the $T$-matrix, the Gaussian form factor $f_{\Lambda} (\vec{p})$ appearing in the two-point loop function $G_{\text{e}}$ should also be included in the $T$-matrix~\cite{Cincioglu:2016fkm,Ye:2025ywy}.
Thus,
the elastic part of $T$-matrix can be rewritten as
\begin{align}
    T_{\text{ee}}'(E)=f_{\Lambda}(\vec{p})\left[\left[\hat{V}_{\text{e}}^{\text{eff}}(E)\right]^{-1}-G_\text{e}(E)\right]^{-1}f_{\Lambda}(\vec{p}\,^\prime)
\end{align}
where $\vec{p}$ and $\vec{p}\,^\prime$ denote the three-momenta of the particles in the c.m. frame of the initial and final two-body systems, respectively. Analogously, the contribution to $T_{\text{ie}}$ can be expressed as
\begin{align}
    T_{\text{ie}}'(E)& = T_{\text{ie}}f_{\Lambda}(\vec{p}\,^\prime)\notag\\
    & = V_{\text{in}} f_{\Lambda}(\vec{p}\,^\prime)+V_{\text{in}}G_{\text{e}}(E)f_{\Lambda}(\vec{p})^{-1}T_{\text{ee}}^\prime(E).
\end{align}
It is worth noting that the form factor can also be introduced to each vertex in the potential, instead in the two-point loop function and external particles. In this way, the unitarity of the $T$-matrix is automatically preserved.

\subsection{The invariant mass distribution}
\begin{figure}
    \centering
    \includegraphics[width=1\linewidth]{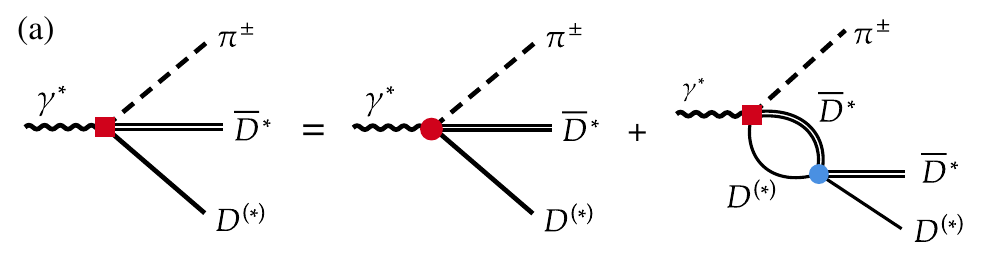}
    \includegraphics[width=1\linewidth]{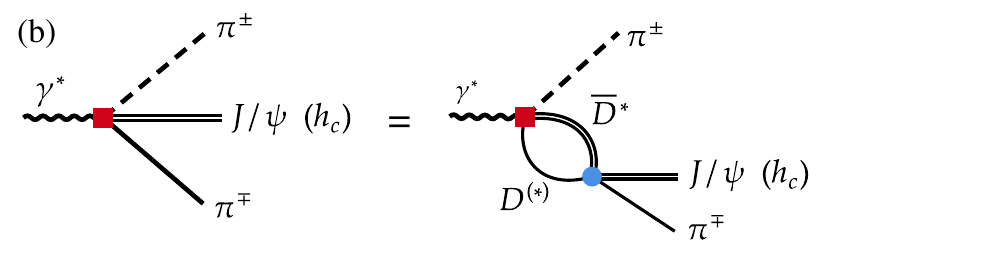}
    \caption{Graphical representation for the decay amplitudes of both elastic channel (a) and inelastic channel (b). The red filled squares and circles indicate the physical and bare decay amplitudes, respectively. 
     The blue filled circle in panel (a) denotes the transitions between the two elastic channels, whereas that in panel (b) corresponds to the transitions between the elastic and inelastic channels.}
    \label{Phy_pro_amp}
\end{figure}

The physical production amplitude can be written as \cite{Zhuang:2021pci}
\begin{align}
    \mathcal{F}(E)=\mathcal{P}+T(E)G(E)\mathcal{P},
    \label{pro_amp}
\end{align}
where $\mathcal{P}=([\mathcal{P}_{\text{e}}]_{2\times 1}^\text{T},\bm{0})^\text{T}$ denotes the bare production amplitude. Here, $\mathcal{P}_\text{e}$ represents the bare production amplitude between the virtual photon and charmed (or bottom) meson pairs. 
The direct production amplitudes for the inelastic channels are expected to be 
suppressed. Hence, the bare production amplitudes of the inelastic channels are assumed to be vanished. Substituting Eqs.~(\ref{T}) and (\ref{G}) into Eq.~(\ref{pro_amp}), one obtains the explicit expression for the physical production amplitude
\begin{align}
    \left(\begin{array}{c}
         \mathcal{F}_{\text{e}}  \\
          \mathcal{F}_{\text{i}}
    \end{array}\right)=\left(\begin{array}{c}
\mathcal{P}_{\text{e}}+T_{\text{ee}}G_{\text{e}}\mathcal{P}_{\text{e}}\\
T_{\text{ie}}G_{\text{e}}\mathcal{P}_{\text{e}}
    \end{array}\right),
    \label{pa}
\end{align}
where $\mathcal{F}_{\text{e}}$ and $\mathcal{F}_{\text{i}}$ correspond to the physical production amplitudes for the elastic and inelastic channels, respectively. The corresponding diagrams are shown in Fig.~\ref{Phy_pro_amp}.
Analogously to the treatment of the $T$-matrix, a Gaussian form factor $f_{\Lambda}(\vec{p})$ is introduced to regularize the loop integral. As a result, the elastic production amplitude should be redefined as
\begin{align}
    \mathcal{F}_{\text{e}}'=f_{\Lambda}(\vec{p})\mathcal{F}_{\text{e}},
    \label{fe}
\end{align}
while the inelastic amplitudes $\mathcal{F}_{\text{i}}$ remain unchanged.

For the elastic channels in the $Z_c$ and $Z_b$ systems, the pion is treated as a spectator. Since the virtual photon carries quantum numbers $J^{PC}=1^{--}$, the relative orbital angular momentum between the pion and the charmed (or bottom) meson pairs must be even. In this work, the pion momentum can reach up to 0.8~GeV, so the contribution for $D$-wave cannot be neglected. As the result, we consider both the $S$-wave and $D$-wave $D^{(*)}\bar{D}^*\pi$ bare production amplitudes. 
The bare production amplitudes for $S$-wave $D\bar{D}^*\pi$ and $D^*\bar{D}^*\pi$ are defined as
\begin{align}
    P(D\bar{D}^*\pi) =  U_1^{\text{c}},\quad
    P(D^*\bar{D}^*\pi) = U_2^{\text{c}}.
\end{align}
The bare production amplitudes for $D$-wave $D\bar{D}^*\pi$ and $D^*\bar{D}^*\pi$ can be defined as
\begin{align}
    P(D\bar{D}^*\pi) =  U_3^{\text{c}}{p_{\pi}^2},\quad P(D^*\bar{D}^*\pi) = U_4^{\text{c}}{p_{\pi}^2},
\end{align}
with $p_\pi$ the three momentum of pion in the $e^+e^-$ c.m. framework. 
Therefore, the explicit form of the bare production amplitude is given by
\begin{align}
    \mathcal{P}_\text{e}=\left(\begin{array}{c}
         P(D\bar{D}^*\pi)  \\
        P(D^*\bar{D}^*\pi)
    \end{array}\right)=\left(\begin{array}{c}
         U_1^{\text{c}}+ U_3^{\text{c}}{p_{\pi}^2} \\
        U_2^{\text{c}}+U_4^{\text{c}}{p_{\pi}^2}
    \end{array}\right),
\end{align}
where the same treatment is applied to the $Z_b$ system.  One should note that the coefficients $U_i$ are energy dependent, and their values vary at different electron–positron collision energies. In our work, electron–positron collision energy of the $Z_b$ system is fixed at 10.866~GeV, so only one set of parameters $U_i^\text{b}$ is
required. In contrast, for the $Z_c$ system, the c.m. energies are 4.23~GeV and 4.26~GeV, and therefore two independent sets of parameters $U_i^{\text{c}}$ are introduced.

The differential decay rate for a three-body decay process is expressed as 
\begin{align}
    \mathrm{d}\Gamma=\frac{1}{(2\pi)^332M^3}|\mathcal{M}(m_{12})|^2dm_{12}^2dm_{23}^2,
\end{align}
where $M$ denotes the mass of the parent particle, and $m_{ij}^2=(p_i+p_j)^2$ represents the invariant mass squared of particles $i$ and $j$. The symbol $\mathcal{M}$ stands for the decay amplitude. In the present analysis, the pion is treated as the spectator (third) particle, and the corresponding decay amplitudes for the elastic channels are given by $\mathcal{F}^\prime_{\text{e}}$ as defined in Eq.~(\ref{fe}).
Consequently, the invariant mass distributions of the elastic channels can be written as 
\begin{eqnarray}
    \frac{\mathrm{d}\Gamma_\text{e}}{\mathrm{d}m_{12}}=\frac{|\mathcal{F}^\prime_\text{e}|^2p_\pi k_e}{32\pi^3 s},
    \label{distr1}
\end{eqnarray}
where $s$ denotes the c.m. energy squared of the $e^+e^-$ system, and $k_\text{e}=\sqrt{\lambda(m_{12}^2,m_1^2,m_2^2)}/(2 m_{12})$ represents the three-momentum of the $D^{(*)}$ (or $B^{(*)}$) meson in the c.m. frame of the corresponding charmed (or bottom) meson pair. Here the standard K\"allen function is given by
\begin{eqnarray}
    \lambda(x,y,z)=x^2+y^2+z^2-2xy-2yz-2xz.
\end{eqnarray}
$m_1$ and $m_2$ are the masses of $D^{(*)}$(or $B^{(*)}$) and $\bar{D}^{(*)}$(or $\bar{B}^{(*)}$) mesons, respectively. The pion momentum in the $e^+e^-$ c.m. frame can be obtained by $ p_\pi=\sqrt{\lambda(s,m_\pi^2,m_{12}^2)}/(2\sqrt{s})$. Similarly, the invariant mass distributions of the inelastic channels take the form 
\begin{eqnarray}
    \frac{\mathrm{d}\Gamma_\text{in}}{\mathrm{d}m_{12}}=\frac{|{\mathcal{F}_\text{in}}|^2p_\pi k_{\text{in}}}{32\pi^3 s},
    \label{distr2}
\end{eqnarray}
where the amplitude $\mathcal{F}_{\text{in}}$ is obtained from Eq.~(\ref{pa}).  $k_{\text{in}}=\sqrt{\lambda(m_{12}^2,m_1^2,m_2^2)}/(2 m_{12})$ denotes the three-momentum of particle $X$ in the $X\pi$ c.m. frame. The set of inelastic channels $X$ includes the states $J/\psi$, $h_c$, $\Upsilon(1S)$, $\Upsilon(2S)$,  $h_b(1P)$ and $h_b(2P)$. The detailed derivations of Eq.~(\ref{distr1}) and Eq.~(\ref{distr2}) are provided in Appendix~\ref{three_body}.

To fit the experimental event distributions of the $Z_c$ system, the fitting function for each channel is rewritten as 
\begin{align}
    \text{event}_i= \mathcal{N}_i\times\frac{d\Gamma_i}{dm_{12}},
\end{align}
where $i$ denotes the $i$-th channel. It is worth noticing that $e^+e^-\to D^0D^{*-}\pi^+$~($\sqrt{s}$=4.26~GeV) channel is considered as the reference channel, whose fitting function is not multiplied by the factor $\mathcal{N}_i$, and $\mathcal{N}_i$ accounts for the relative strength of the $i-$th channel with respective to the $D^0D^{*-}\pi^+$ channel. The same treatment is conducted in the $Z_b$ case with $e^+e^-\to B^0B^{*-}\pi^+$ as the reference one.

%% file: Results.tex
\section{RESULTS AND DISCUSSION}\label{Results}
\begin{figure*}
    \centering
    \includegraphics[width=0.325\linewidth]{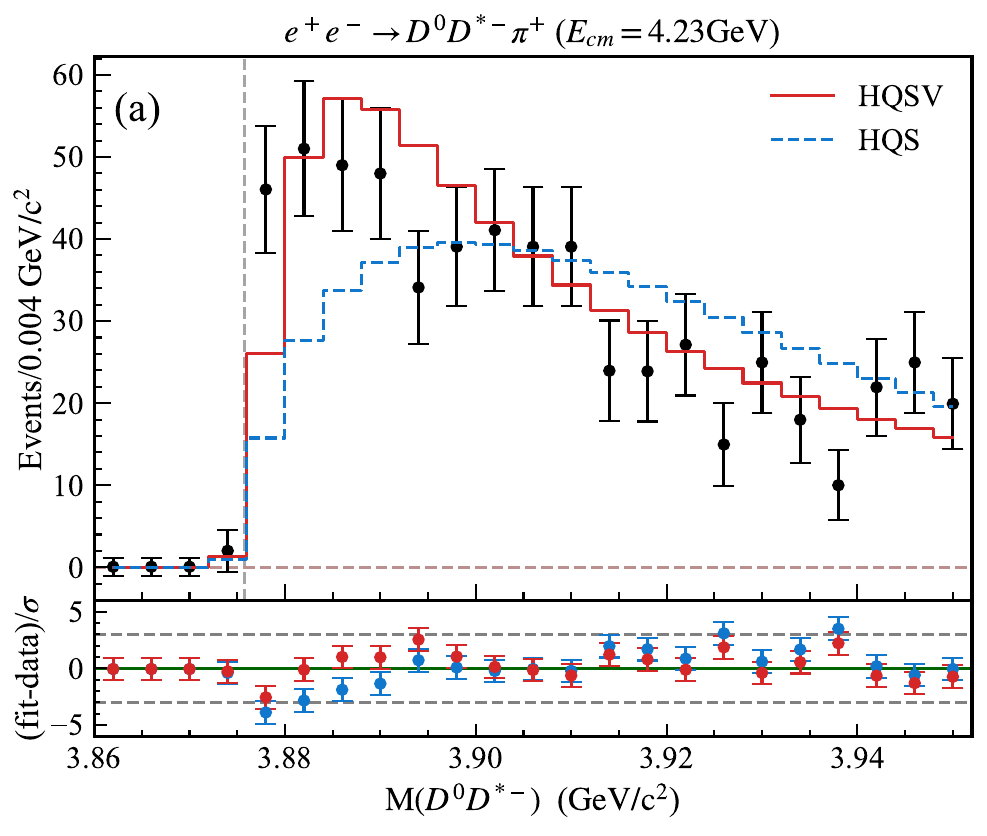}
    \includegraphics[width=0.325\linewidth]{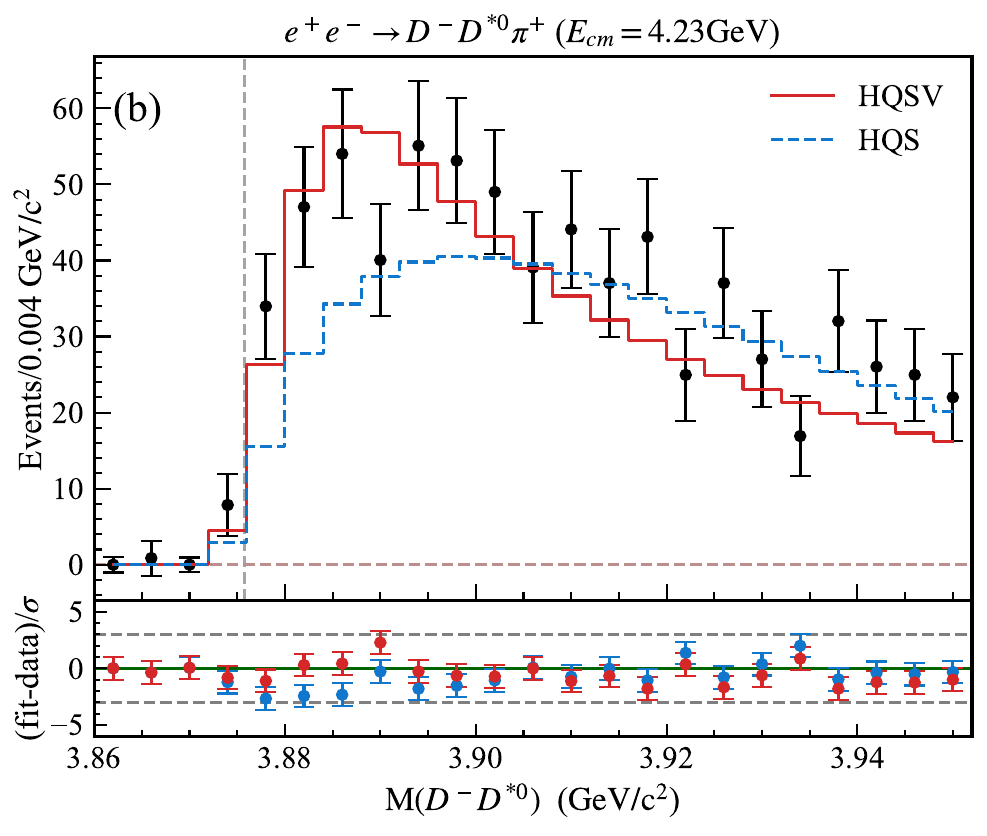}
    \includegraphics[width=0.325\linewidth]{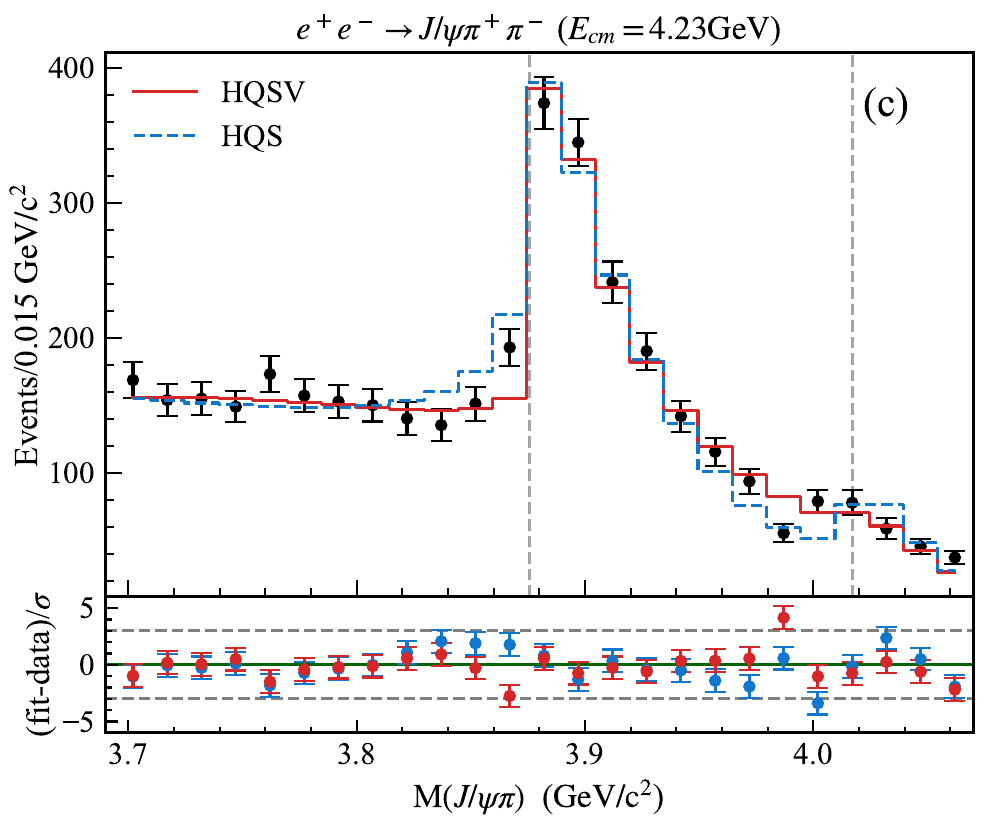}
    \includegraphics[width=0.325\linewidth]{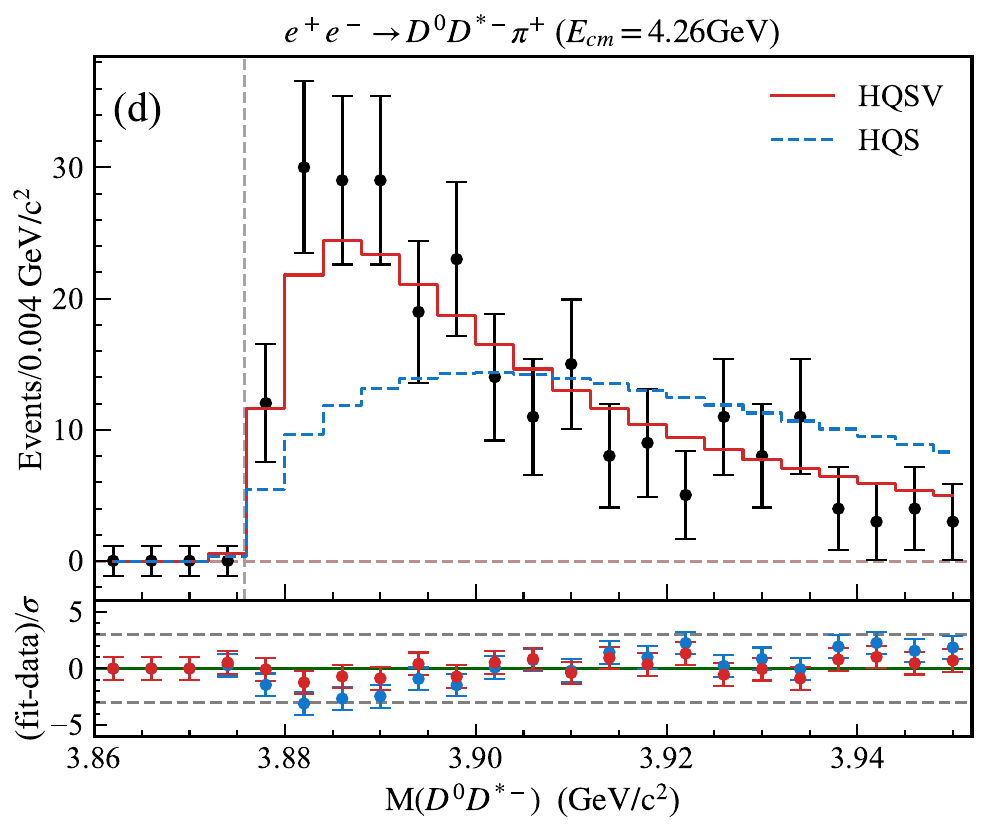}
    \includegraphics[width=0.325\linewidth]{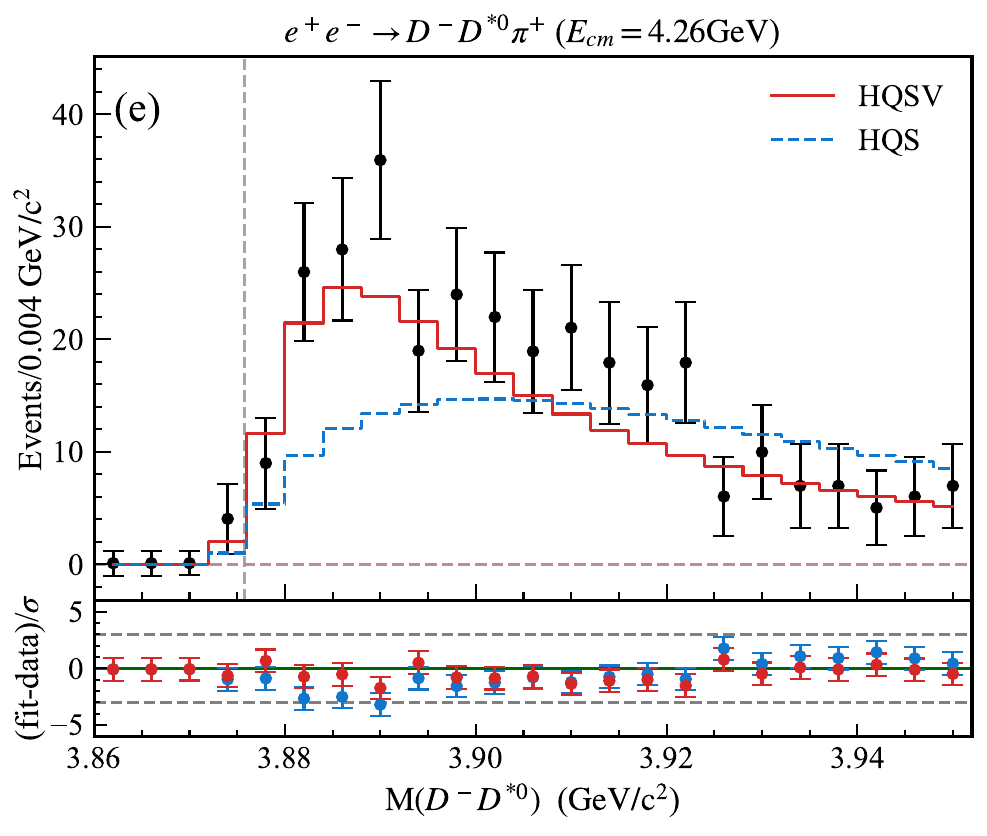}
    \includegraphics[width=0.325\linewidth]{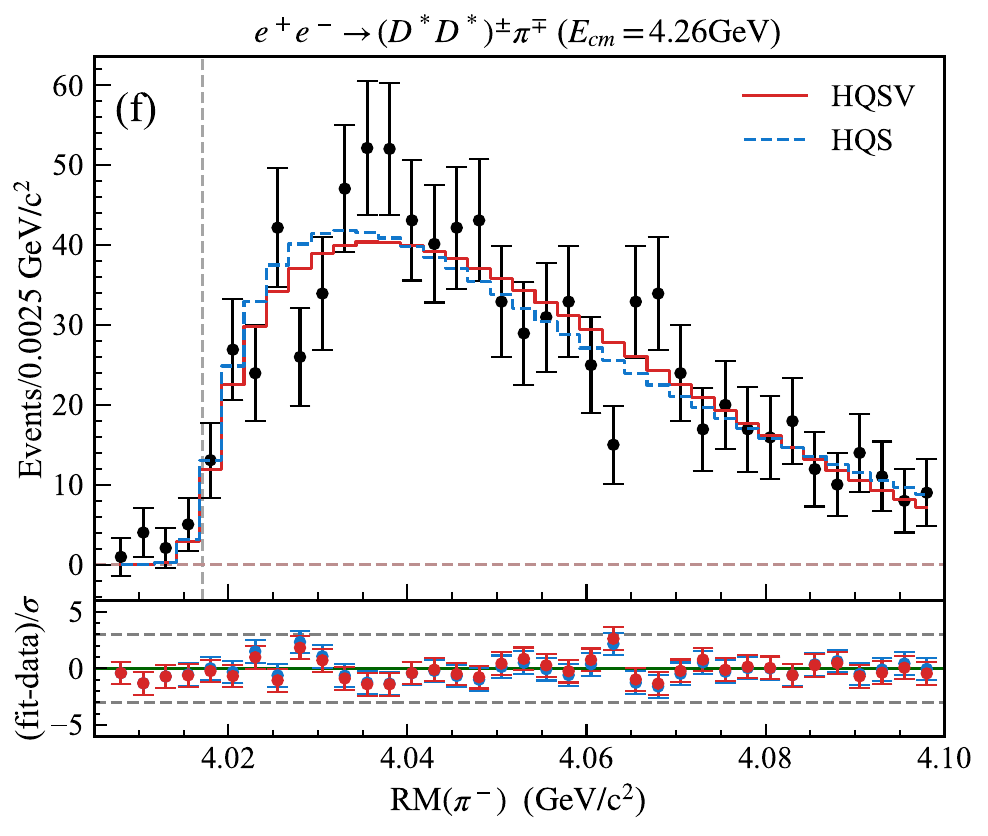}
    \includegraphics[width=0.325\linewidth]{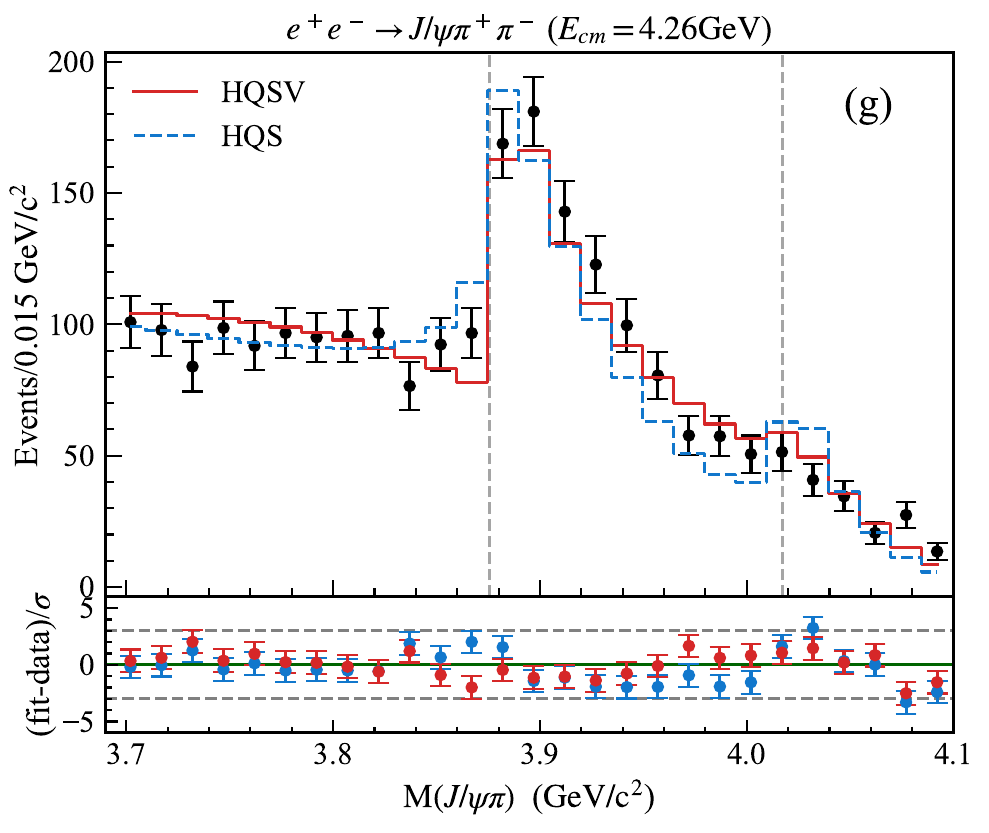}
    \includegraphics[width=0.325\linewidth]{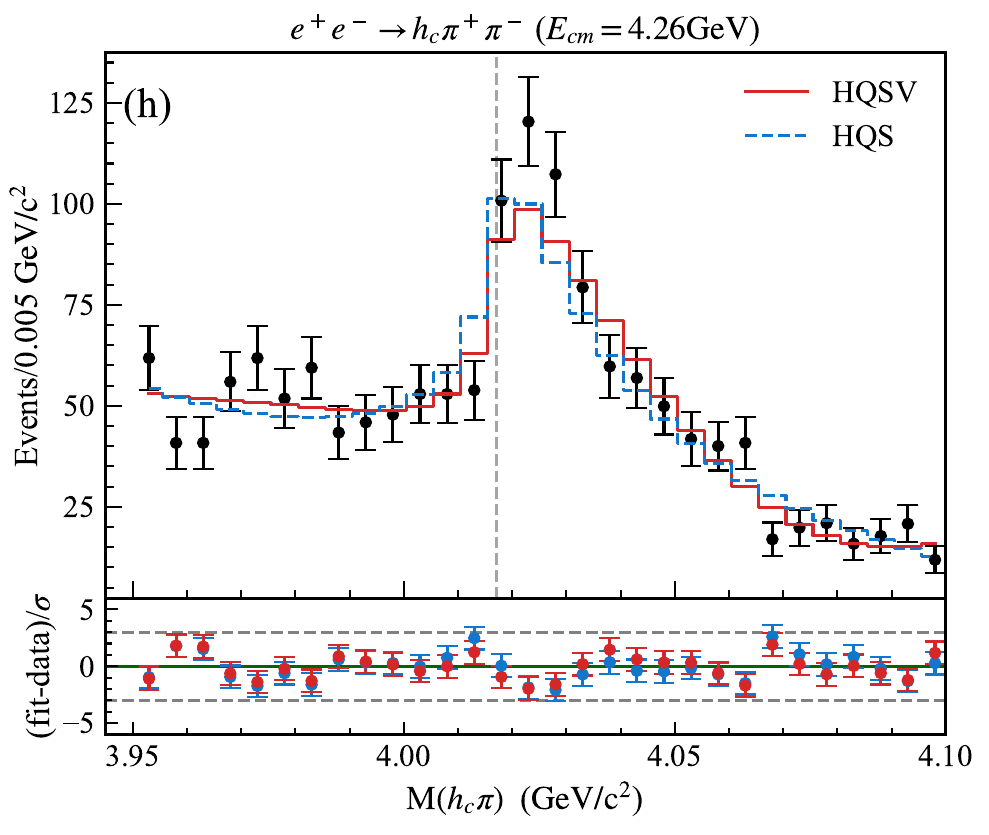}
    \caption{Fitted line shapes of both HQSV (red solid curves) and HQS (blue dashed curves) schemes for the charm system in comparison with the experimental data~\cite{BESIII:2013mhi,BESIII:2015pqw,BESIII:2017bua} from BESIII collaboration. Subfigures (a)–(c) correspond to the invariant-mass distributions at $\sqrt{s}=4.23$~GeV, while subfigures (d)-(h) show those at $\sqrt{s}=4.26$~GeV. The vertical two gray dashed lines indicate the $D\bar{D}^*$ and $D^*\bar{D}^*$ thresholds from left to right, in order. The lower panel in each subfigure presents the standardized residuals, where the red and blue points correspond to the HQSV and HQS cases, respectively.}
    \label{line_shape_Zc}
\end{figure*}
\begin{figure*}
    \centering
    \includegraphics[width=0.325\linewidth]{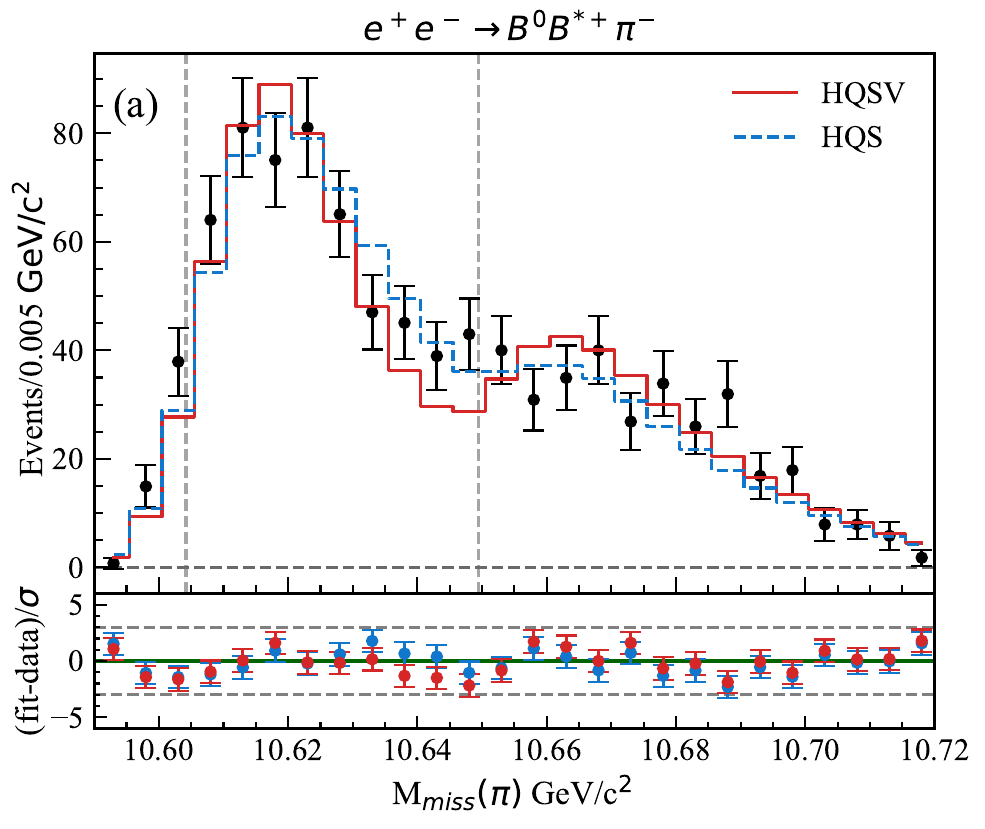}
    \includegraphics[width=0.325\linewidth]{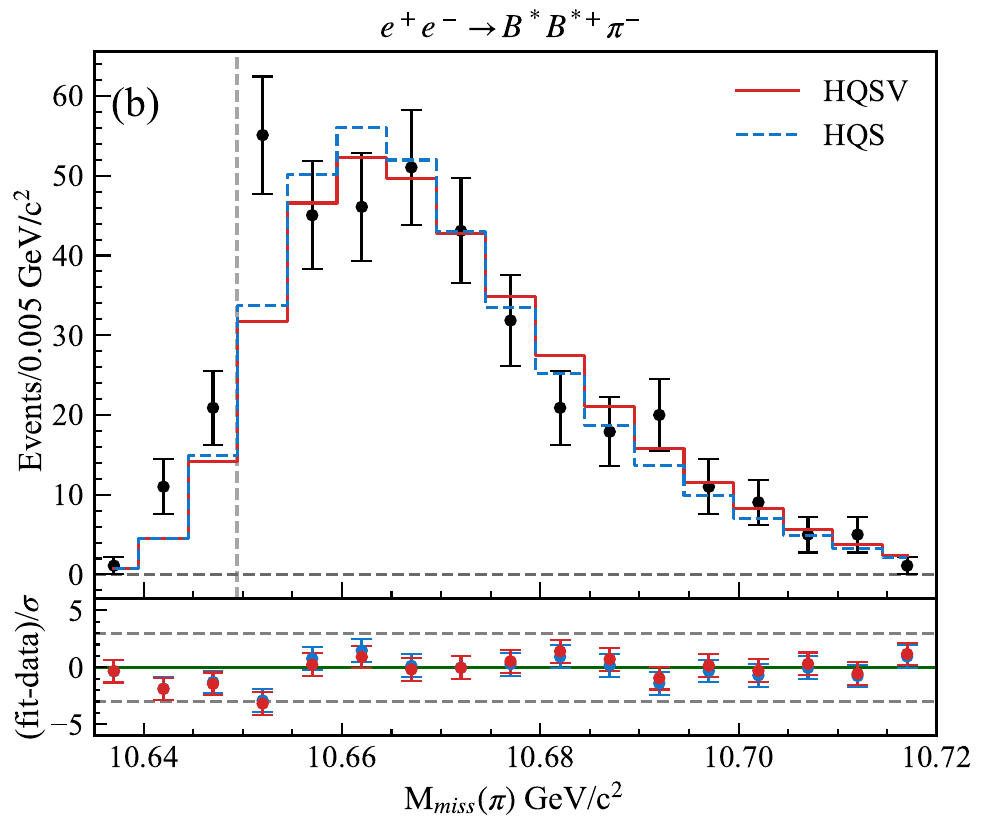}
    \includegraphics[width=0.325\linewidth]{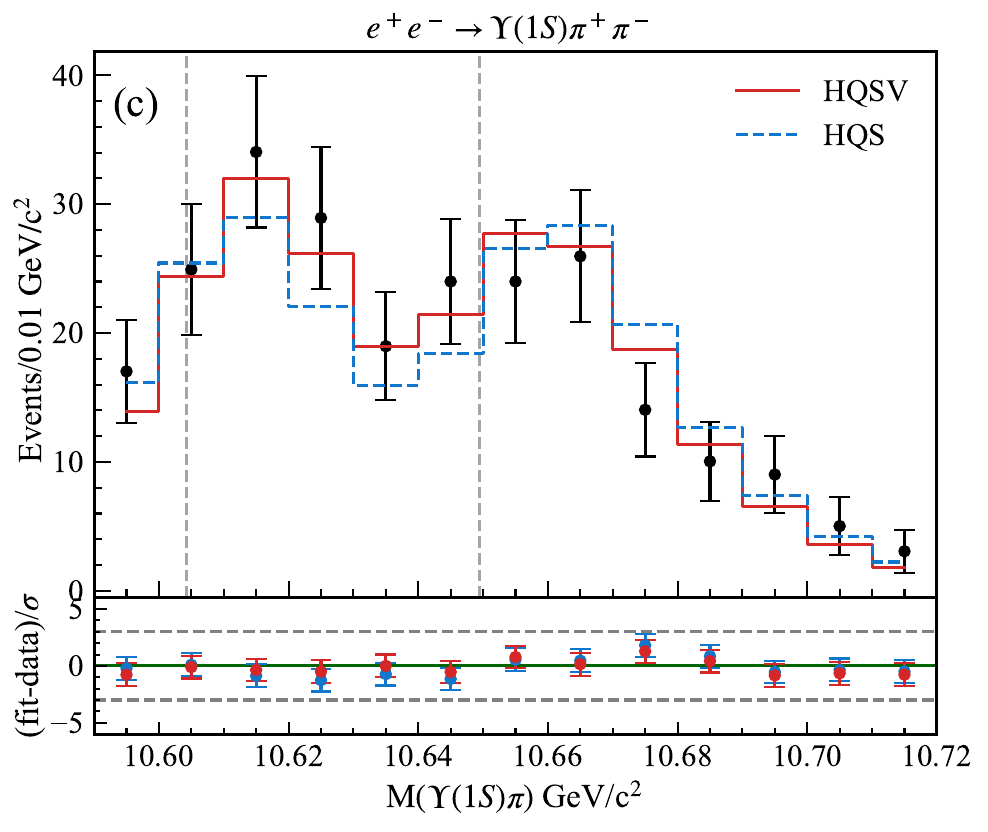}
    \includegraphics[width=0.325\linewidth]{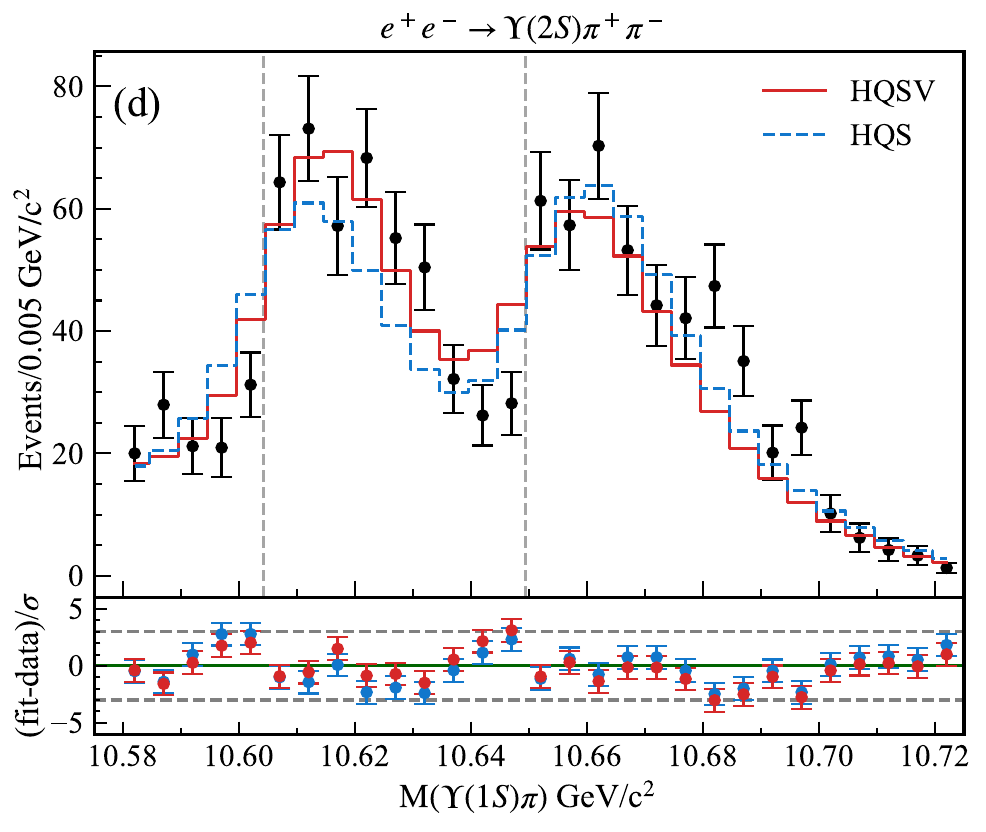}
    \includegraphics[width=0.325\linewidth]{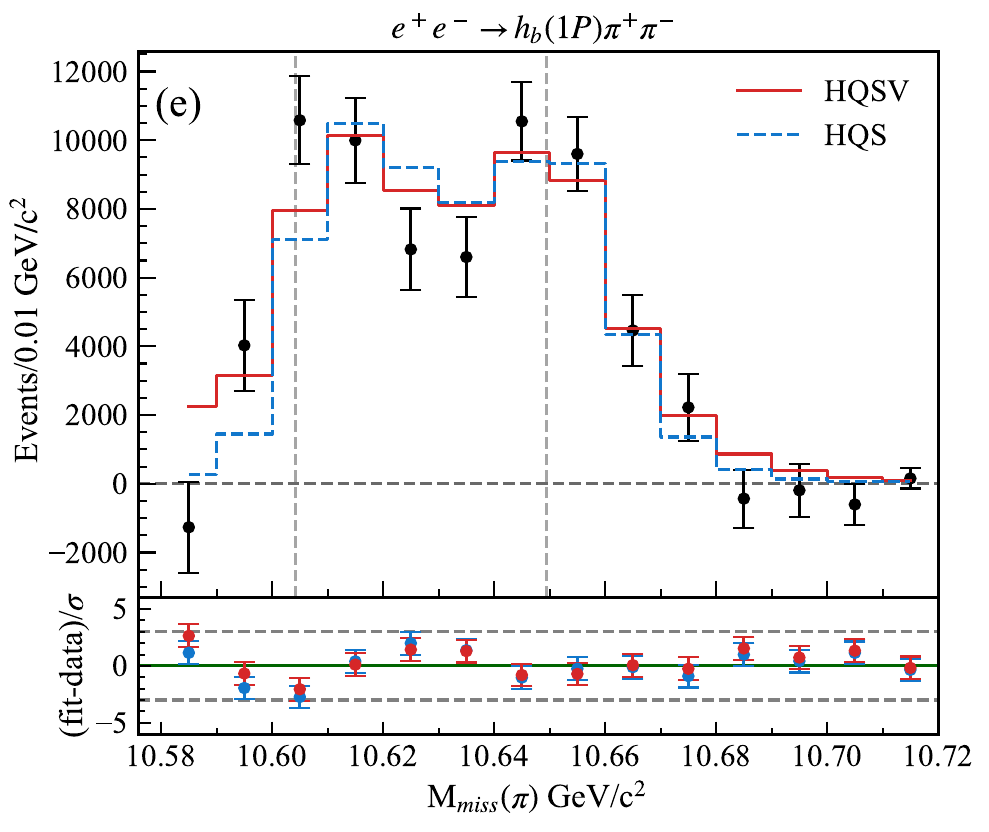}
    \includegraphics[width=0.325\linewidth]{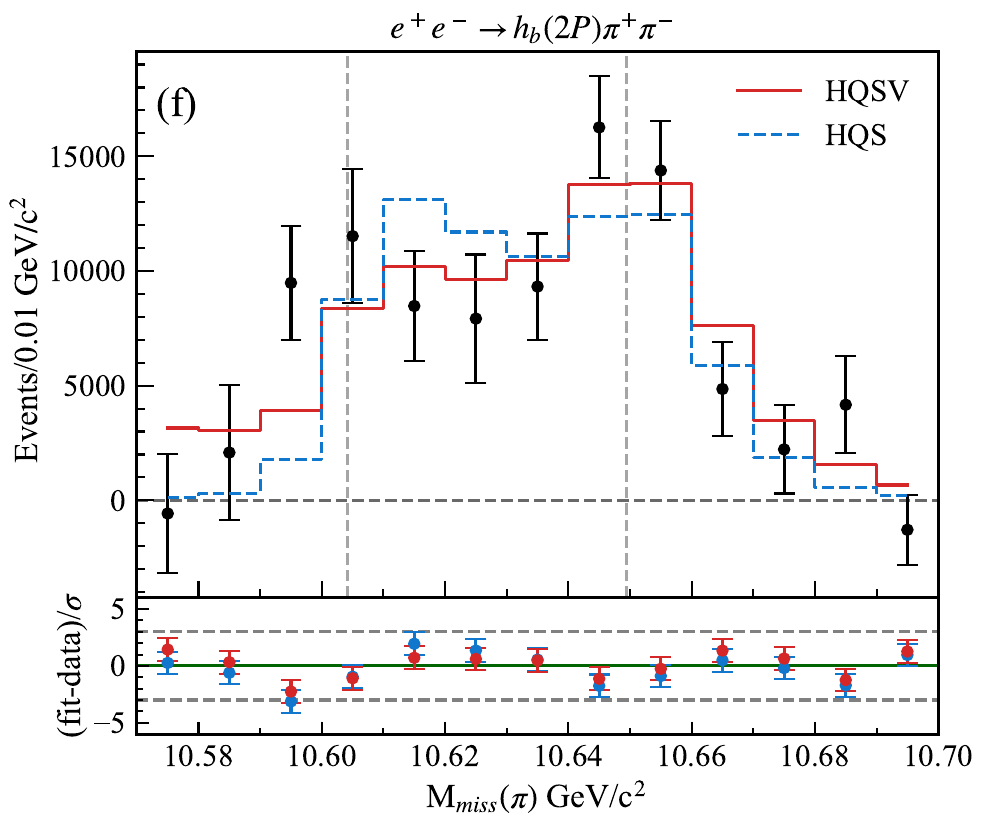}
    \caption{Fitted line shapes of both HQSV (red solid curves) and HQS (blue dashed curves) schemes for the bottom system in comparison with the experimental data~\cite{Belle:2011aa,Belle:2014vzn,Belle:2015upu} from Belle collaboration. The two vertical  gray dashed lines indicate the $B\bar{B}^*$ and $B^*\bar{B}^*$ thresholds from left to right, in order. The lower panels are depicted in the same manner as those in Fig.~\ref{line_shape_Zc}.}
    \label{line_shape_Zb}
\end{figure*}
In this section, we perform a global fit to the invariant-mass distributions of subsystems at various c.m. energies. Two fitting schemes are considered. In Scheme-I, illustrated by the charm sector, the symmetry-breaking contributions in both elastic and inelastic channels are neglected, i.e. the parameters $\mathcal{B}$, $v_{1S}^{\psi}$ and $v_{1P}^{h_c}$ are fixed to zero, which we refer to as HQS scheme. In Scheme-II, the symmetry-breaking effects are taken into account by allowing $\mathcal{B}$, $v_{1S}^{\psi}$ and $v_{1P}^{h_c}$ to vary in the fit, and this scheme is denoted as HQSV. The same procedure is applied to the bottom sector. The fitting is performed by imimuit~\cite{JAMES1975343} with more than 10000 randomized initial values to ensure convergence to a global minimum. The best fitted line shapes of the invariant-mass distributions for the $Z_c$ and $Z_b$ systems are shown in Fig.~\ref{line_shape_Zc} and Fig.~\ref{line_shape_Zb}, respectively. 
The lower panels of both figures illustrate that the  
standardized residual distributions, which indicate that the majority of data points fall within the interval $[-3,3]$ for all the fitting schemes. The projections of these residuals onto the vertical axis are 
displayed in Appendix~\ref{residual}.  
\begin{table*}
\renewcommand{\arraystretch}{1.5}
    \centering
    \caption{The fitted parameters of the $Z_c$ and $Z_b$ systems. $\Lambda$ is the parameter in the Gaussian form factor.  $C_0$, $C_1$, and $\mathcal{B}$ correspond to the interaction potentials between the elastic channels, while the remaining parameters describe those between elastic and inelastic channels.}
    \label{para}
    \begin{tabular}{p{3cm}<{\centering}|p{2.5cm}<{\centering}p{2.5cm}<{\centering}|p{2.5cm}<{\centering}p{2.5cm}<{\centering}}
    \hline\hline
   \textbf{Parameters}  &  \textbf{HQSV}~$(Z_c)$ & \textbf{HQS}~$(Z_c)$ & \textbf{HQSV}~$(Z_b)$ & \textbf{HQS}~$(Z_b)$  \\
    \hline
   $C_0~[\mathrm{GeV^{-2}}] $  &  $-3.88\pm 0.37$  &$-6.75\pm 0.13$& $-2.21\pm 0.06$ & $-2.64\pm 0.12$  \\
   $C_1~[\mathrm{GeV^{-2}}]$  &  $9.41\pm 1.00$ &$-3.32\pm 0.29$&  $-2.67 \pm 0.07$   & $-2.88\pm 0.11$\\
   $\mathcal{B}~[\mathrm{GeV^{-2}}]$  & $-14.65\pm 0.46$ &$-$& $-0.25\pm 0.04$ &$-$  \\ 
   $g_{1S}^{\psi}~[\mathrm{GeV^{-3/2}}]$  & $-5.34\pm 0.09$ &$-2.15\pm 0.07$& $-$ & $-$\\
   $g_{1P}^{h_c}~[\mathrm{GeV^{-2}}]$  &  $0.98\pm 0.06$ &$-0.31\pm 0.04$& $-$ & $-$ \\
   $v_{1S}^{\psi}~[\mathrm{GeV^{-3/2}}]$  & $1.99\pm 0.12$ &$-$&  $-$ & $-$ \\
   $v_{1P}^{\psi}~[\mathrm{GeV^{-2}}]$  & $0.27\pm 0.02$ &$-$&  $-$ & $-$ \\
   $g_{1S}^{\Upsilon}~[\mathrm{GeV^{-3/2}}]$  & $-$ & $-$& $-1.42\pm 0.05$ & $-0.47\pm 0.12$ \\
   $g_{2S}^{\Upsilon}~[\mathrm{GeV^{-3/2}}]$  & $-$  &$-$  & $-0.13\pm 0.02$ & $-1.99\pm 0.08$ \\
   $g_{1P}^{h_b}~[\mathrm{GeV^{-2}}]$  & $-$  & $-$ & $0.15 \pm 0.04$ & $0.24\pm 0.06$ \\
   $g_{2P}^{h_b}~[\mathrm{GeV^{-2}}]$  & $-$  & $-$ & $-4.30 \pm 0.51$ & $-4.95\pm 0.64$ \\
   $v_{1S}^{\Upsilon}~[\mathrm{GeV^{-3/2}}]$  & $-$ & $-$& $-0.26\pm 0.07$ & $-$ \\
   $v_{2S}^{\Upsilon}~[\mathrm{GeV^{-3/2}}]$  & $-$  & $-$ & $-0.03\pm 0.01$ & $-$ \\
   $v_{1P}^{h_b}~[\mathrm{GeV^{-2}}]$  & $-$  & $-$ & $0.03 \pm 0.01$ &$-$  \\
   $v_{2P}^{h_b}~[\mathrm{GeV^{-2}}]$  & $-$  &$-$  & $-0.50 \pm 0.28$ & $-$ \\
   $\Lambda~[\mathrm{GeV}]$  & $1.00$  &$1.80$  & $1.80$ & $1.40$ \\\hline
   $\chi^2/\mathrm{d.o.f.}$ & 1.12  & 1.92 & 1.76 & 1.81 \\
   \hline
   \hline
   \end{tabular}  
\end{table*}

In the bottom sector, the reduced chi-squares are $\chi^2/\mathrm{d.o.f.}=1.76$ and $\chi^2/\mathrm{d.o.f.}=1.81$ for the HQSV and HQS schemes, respectively. One can see that the HQS breaking effect does not substantially improve the quality of the fit,  
which indicates that the HQSS already works well in the bottom sector as we expected. In the charm sector,   
 the reduced chi-squares are $\chi^2/\mathrm{d.o.f.}=1.12$ and $\chi^2/\mathrm{d.o.f.}=1.92$ for the HQSV and HQS schemes, respectively. The HQS breaking effect significantly improve the fit quality, as the charm quark mass is not as heavy as the bottom quark mass.
To further assess the robustness of these results, we examine the stability of our model against variations of the form factor. Specifically, the Gaussian form factor in the two-point function of Eq.~(\ref{G_fun}) is replaced with a monopole form factor, and the fits are repeated. As a preliminary test, we restrict the analysis to the HQSV scheme. The refitted line shapes 
 and the pole positions show only minor deviations from those obtained with the Gaussian form factor, in both the $Z_c$ and $Z_b$ systems. This consistency indicates that the results are relatively insensitive to the choice of the form factor, demonstrating good model stability and a weak model dependence. More details can be found in Appendix~\ref{model_d}.
Consequently, in the following sections, we present and discuss the results for the Gaussian-type form factor, without loss of generality of the model.
The corresponding dynamical parameters  that govern the scattering amplitudes are listed in Tab.~\ref{para} of Appendix~\ref{FIT}. From those parameters, 
one can extract the physical information of interest, such as the pole positions in the complex energy plane, the effective coupling constants and so on, which will be discussed one by one. 

\subsection{Pole analysis}

A state corresponds to a pole of the $T$-matrix in the complex energy plane, which may appear as a bound state, a virtual state, or a resonance. The pole positions are determined by solving the equation
\begin{align}
    \textbf{det}\left[1-\hat{V}_{\text{ee}}^{\text{eff}}G_{\text{e}}(E_r)\right]=0~.
\end{align}
By analytic continuation, the complex $E$-plane can be mapped to $2^n$ Riemann sheets (RSs), with $n$ corresponding to the number of coupled channels. Each sheet is specified by a label of the form $(\pm,\dots,\pm)$, which reflects the signs of the imaginary parts of the c.m. three-momenta in the respective two-body channels. The physical sheet is denoted as $(+,\dots,+)$, while others correspond to unphysical ones. The physical and unphysical RSs are connected, i.e., 
\begin{align}
    G_{ii}^{\text{II}}(E-i\varepsilon)=G_{ii}^{\text{I}}(E+i\varepsilon),
\end{align}
above the corresponding branch cut. 
Here, I and II indicate the two-point functions on the physical and unphysical RSs, respectively, and $i=1,\dots,n$ denote the $i$-th channel. 

In general, the search for poles requires examining all the $2^n$ RSs. However, in practice, only the poles located on the physical sheet $(+,+,\dots,+)$ and on the unphysical sheets directly connected to the physical region are of physical significance. The unphysical RSs are conventionally labeled by sequentially replacing the plus signs with minus signs, i.e. $(-,+,\dots,+)$, $(-,-,\dots,+)$,$\dots$, $(-,-,\dots,-)$. In the $Z_c$ system, four coupled channels are involved, consisting of two elastic and two inelastic ones, with the inelastic thresholds lying below those of the elastic channels. Since our focus is on the pole structures of $Z_c(3900)$ and $Z_c(4020)$, which suffices to search for poles on the two relevant unphysical RSs, namely $(-,-,-,+)$ and $(-,-,-,-)$. On these sheets, the imaginary parts of the c.m. three-momenta of the two inelastic channels always carry a negative sign, corresponding to unphysical propagators. Consequently, the classification of RSs can be reduced to two indices that specify the branch choices of the elastic two-body propagators, labeled as  $(-,+)$ and $(-,-)$. The analytic continuation between these sheets is determined by the threshold structure of the elastic channels. Specifically, in the complex energy plane, the lower half-plane of the $(-,+)$ sheet connects to the upper half-plane of the $(+,+)$ sheet along the real axis in the interval $[\text{thr}_{c1},\text{thr}_{c2}]$, where $\text{thr}_{c1}$ and $\text{thr}_{c2}$ represent the $D\bar{D}^*$ and $D^*\bar{D}^*$ thresholds, respectively. Likewise, the lower half-plane of the $(-,-)$ sheet connects to the upper half-plane of the $(+,+)$ sheet along the real axis in the interval $[\text{thr}_{c2},+\infty)$. Thus, the poles located on the $(-,+)$ and $(-,-)$ RSs influence physical observables only if they lie within the intervals $[\text{thr}_{c1},\text{thr}_{c2}]$ and $[\text{thr}_{c2},E_{\text{max}})$, respectively, where $E_{\text{max}}$ denotes the upper limit of the experimental energy range. Similar analysis also applies to the $Z_b$ case.

In addition to the pole positions, the effective couplings (collected in Tab.~\ref{z_pole}) 
can be extracted at 
a pole at $E_r=M_r-i\Gamma_r/2$, with $M_r$ and $\Gamma_r$ denoting the mass and width of a state, from the $T$-matrix ~\cite{Cincioglu:2016fkm}
\begin{align}
    T_{ij} \sim \frac{g_ig_j}{E-E_r},
\end{align}
where $g_i$ represents the effective coupling of a state to the $i$-th channel. Since $g_i$ is generally complex, its modulus is conventionally adopted to characterize the strength of the coupling. The effective couplings can be extracted from the residues of the $T$-matrix through
\begin{align}
g_i g_i = \lim_{E\to E_r^+}(E-E_r)T_{ii}(E).
\end{align}
\begin{table}
\renewcommand{\arraystretch}{1.5}
    \centering
    \caption{Pole positions and effective couplings to elastic channels in the $Z_c$ and $Z_b$ systems from the best-fit parameters. The units of mass and width are GeV, and the coupling constants have the dimension of $[\mathrm{GeV}]^{-1}$.}
    \label{z_pole}
    \begin{tabular}{p{2.5cm}<{\centering}p{2.5cm}<{\centering}p{2.5cm}<{\centering}}
    \hline\hline
   \textbf{R.S.}$(Z_c)$&  \textbf{HQSV}$(Z_c)$ & \textbf{HQS}$(Z_c)$  \\
    \hline
   $(-,+)$  &  $3.875-0.011i$  & $3.815-0.052i$\\
   $g_{D\bar{D}^*}$  &  0.85  &  1.22\\
   $g_{D^*\bar{D}^*}$  &  0.54  & 0.64\\
   \hline
   $(-,-)$  &  $-$  & $3.988-0.028i$\\
   $g_{D\bar{D}^*}$  & $-$   &0.57 \\
   $g_{D^*\bar{D}^*}$  & $-$   & 1.02\\
   \hline
    \textbf{R.S.}$(Z_b)$&  \textbf{HQSV}$(Z_b)$ & \textbf{HQS}$(Z_b)$  \\
    \hline
   $(-,+)$  &  $10.610-0.010i$  & $10.600-0.013i$\\
   $g_{B\bar{B}^*}$  &  0.42  & 0.39\\
   $g_{B^*\bar{B}^*}$  &  0.11  & 0.13\\
   \hline
   $(-,-)$  &  $10.640-0.021i$  & $10.649-0.016i$\\
   $g_{B\bar{B}^*}$  &  0.25  & 0.13\\
   $g_{B^*\bar{B}^*}$  &  0.42  & 0.40\\
   \hline
   \hline
   \end{tabular}  
\end{table}

The poles obtained on various RSs are collected in Tab.~\ref{z_pole}. For the $Z_b$ system, in both HQS and HQSV schemes, there always exist two poles close to the $B\bar{B}^*$ and $B^*\bar{B}^*$ thresholds, classified as the $Z_b(10610)$ and $Z_b(10650)$. For instance, in the HQS scheme, the two poles are located on the $(-,+)$ and $(-,-)$ sheets, close to the $B\bar{B}^*$ and $B^*\bar{B}^*$ thresholds, respectively. As a result, they are very close to the physical axis and will have significant impact on the physical observables, denoted $Z_b(10610)$ and $Z_b(10650)$, respectively. 
They are consistent with the PDG values of the $Z_b(10610)$ and $Z_b(10650)$, i.e., $10607_{-2.0}^{+2.0}-18.4_{-2.4}^{+2.4}i~\mathrm{MeV}$ and $10652.2_{-1.5}^{+1.5}-11.5_{-2.2}^{+2.2}i~\mathrm{MeV}$,  respectively. 
For the HQSV scheme, there are also two poles on the same Riemann sheet, but with a small deviation from those of the HQS scheme. The pole of the lower $Z_b$ is consistent with the Particle Data Group (PDG) value, but that of the higher one exhibits a moderate deviation from the PDG value.

For the $Z_c$ system, the fit quality of the HQS scheme is relatively poor, indicating that the HQSS framework does not work very well for the charm system. Thus, the corresponding pole position is not reliable. Instead, the HQSV scheme works well for the $Z_c$ system. In this scheme, only a single pole is found on the $(-,+)$ sheet around the $D\bar{D}^*$ threshold, which is identified 
as the $Z_c(3900)$ state. Its extracted mass and width are slightly lower than the values $3887.1_{-2.6}^{+2.6}-28.4_{-2.6}^{+2.6}i$~MeV reported by the PDG, which is extracted from the Breit-Wigner formula.  No pole corresponding to the $Z_c(4020)$ state is reported. The absence of the $Z_c(4020)$ can be easily seen  from the very large value of HQS breaking parameters in Tab.~\ref{para}. We suggest that the $Z_c(4020)$ may originate from a threshold cups effect similar to that in Ref.~\cite{Swanson:2014tra}, given its proximity to the $D^*\bar{D}^*$ threshold, although alternative dynamical mechanisms cannot be ruled out. 

\begin{figure*}
    \centering
    \includegraphics[width=0.3\linewidth]{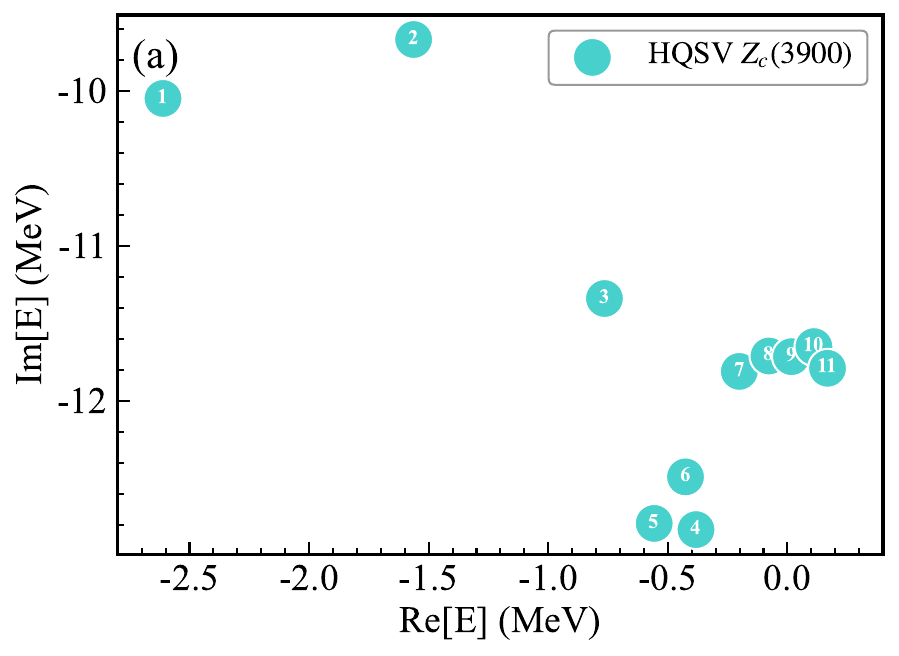}
    \includegraphics[width=0.3\linewidth]{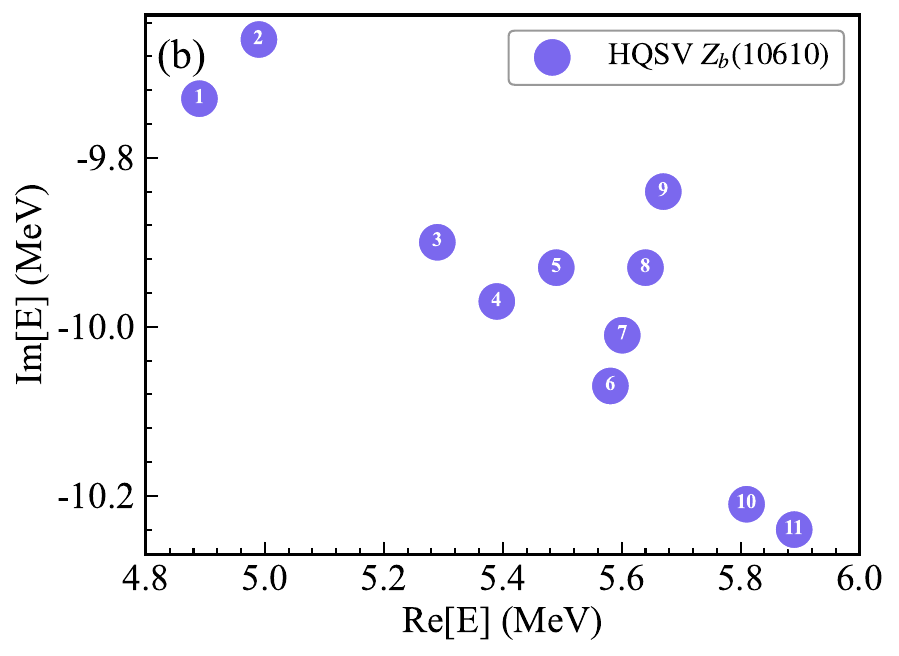}
    \includegraphics[width=0.3\linewidth]{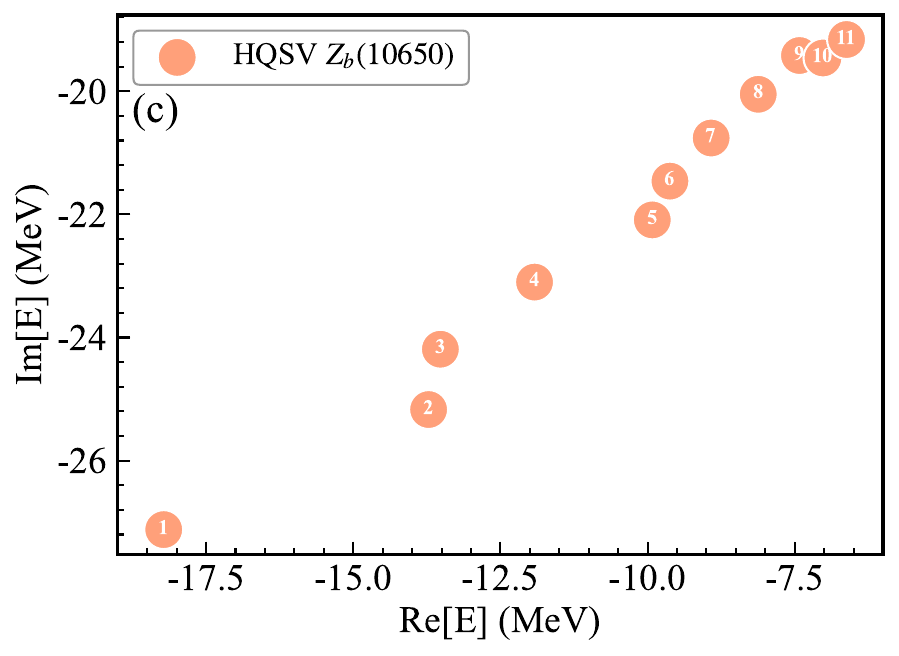}
    \caption{Pole positions in the $Z_c$ and $Z_b$ systems under the HQSV 
    scheme for different cutoff values. The real parts of the poles in panels (a), (b), and (c) are shown relative to the thresholds of $D\bar{D}^*$, $B\bar{B}^*$ and $B^*\bar{B}^*$, respectively. (a) For the $Z_c$ system, the cutoff varies from 0.8 to 1.8 GeV in increments of $0.1~\mathrm{GeV}$, corresponding to eleven distinct values labeled from 1 to 11. (b)-(c) For the $Z_b$ system, the cutoff ranges are from 1.3 to 2.3 GeV with a step of 0.1 GeV, labeled by indices from 1 to 11.}
    \label{lamv}
\end{figure*}

Poles are physical observables, which should be cutoff independence, mainly driven by the experimental data.  
Thus, we refit the experimental data at various cutoffs. For the $Z_c$ system, the cutoff varies from 0.8 to 1.8 GeV with increments of 0.1 GeV. For Fig.~\ref{lamv}(a), we can see that the poles with eleven cutoffs locate within a small energy region, which demonstrates that our framework is stable on top of next leading order corrections. For the $Z_b$ system, the cutoff varies from 1.3 to 2.3 GeV with increments of 0.1 GeV. Fig.~\ref{lamv}(b) (HQS) and (c) (HQSV) show that both the two $Z_b$ poles locate within a small energy region, close to the $B\bar{B}^*$ and $B^*\bar{B}^*$ thresholds, respectively, and exhibit mild variations with respect to the cutoff. 

\begin{figure}
    \centering
    \includegraphics[width=0.475\linewidth]{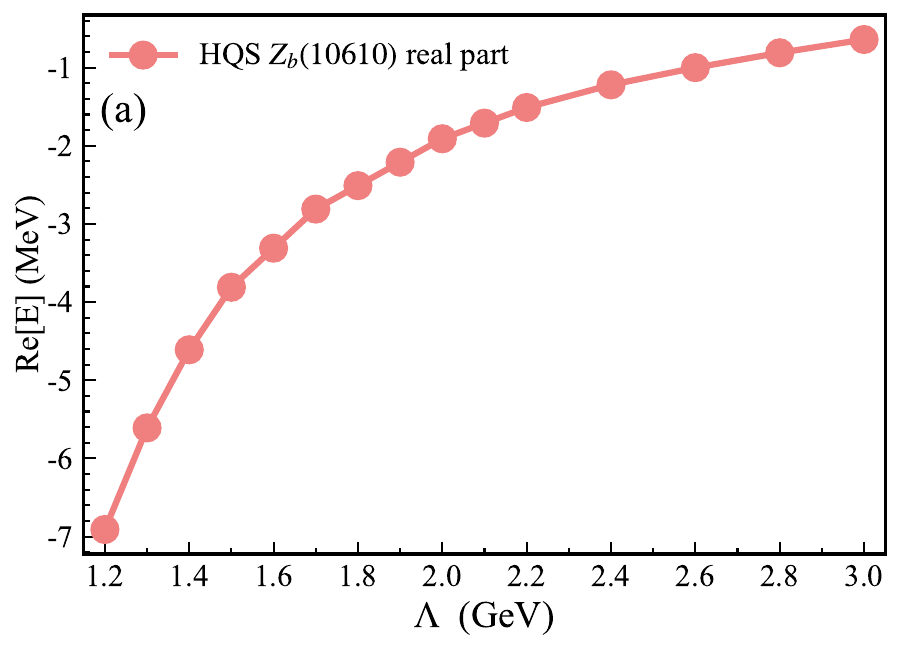}
    \includegraphics[width=0.475\linewidth]{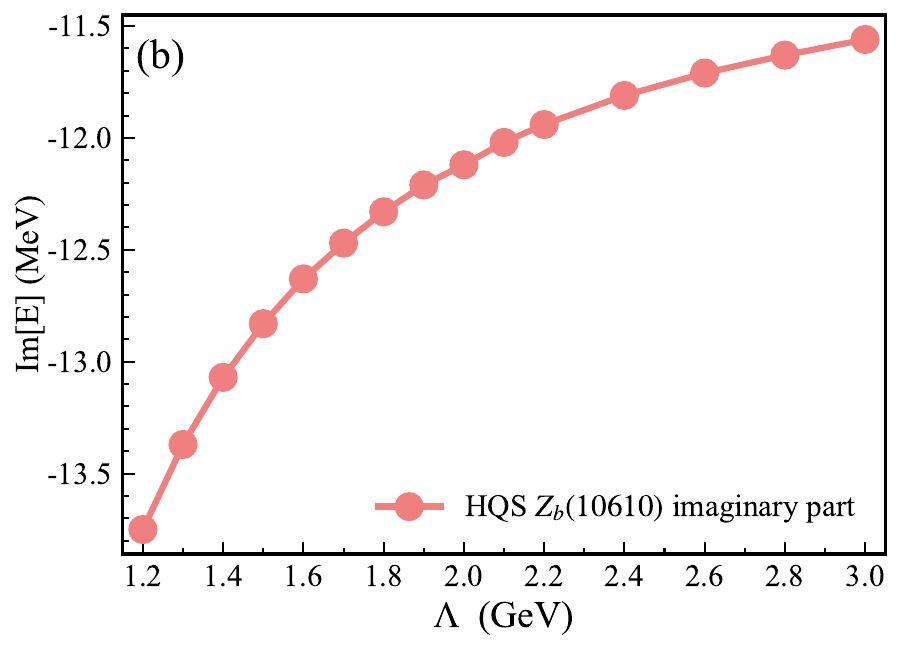}
    \includegraphics[width=0.475\linewidth]{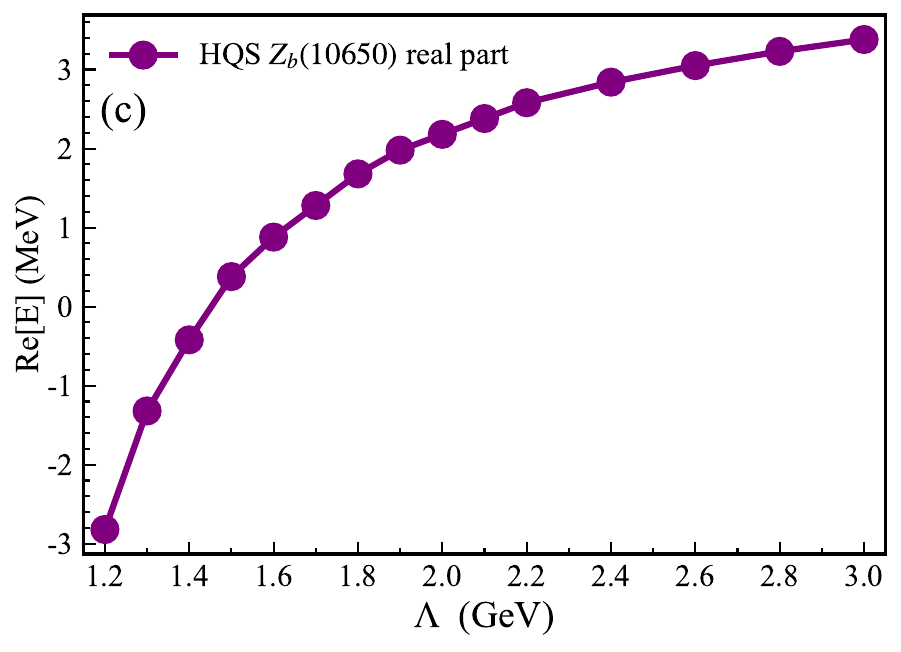}
    \includegraphics[width=0.475\linewidth]{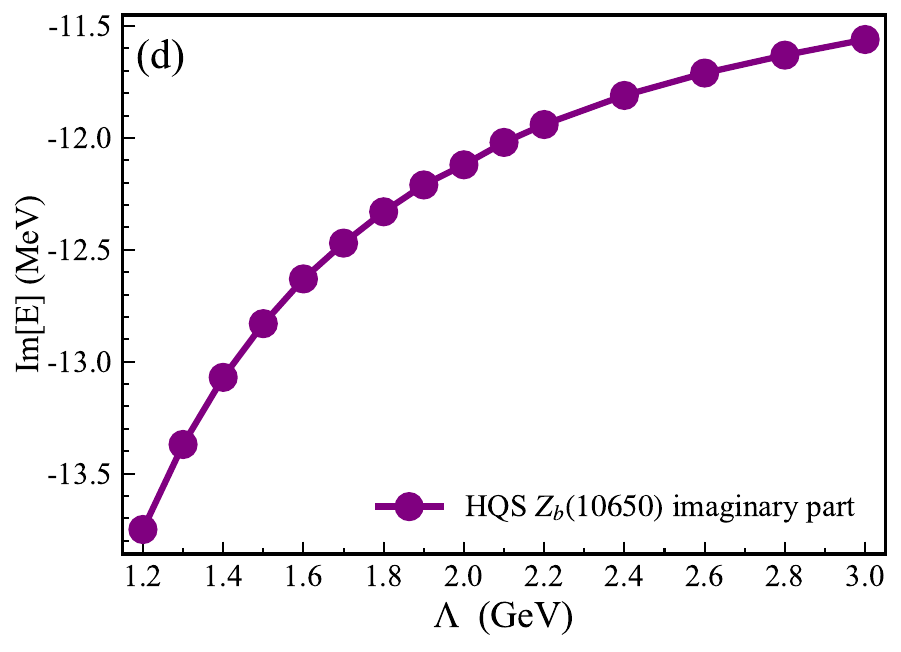}
    \caption{Pole positions in the $Z_b$ system under the HQS scheme for different cutoff values. The real parts of the poles in panels (a) and (c) are shown relative to the   $B\bar{B}^*$ and $B^*\bar{B}^*$ thresholds, respectively. Figures (a) and (b) denote the real and imaginary parts of the $Z_b(10610)$ pole, while (c) and (d) 
    depict those of the $Z_b(10650)$ pole. The cutoff varies from 1.2 to 3.0 GeV, with a step of 0.1 GeV (0.2 GeV) in the region $1.2-2.2$~GeV ($2.2-3.0$ GeV).}
    \label{lams}
\end{figure}

The cutoff dependence can also be used to demonstrate within which scale, HQSS works well for the $Z_b$ system. Thus we plot both the imaginary and real parts of the two poles in terms of the cutoffs in Fig.~\ref{lams}.  
From the figures, one can see that both the imaginary parts and real parts of the two $Z_b$ states exhibit the similar behavior. That is because the values of the two parameters $C_0$ and $C_1$ are of a similar value in HQS scheme. For the HQSV scheme, besides that, the HQS breaking parameters are small in comparison with the HQS conserved parameters.


\subsection{Heavy Quark symmetry breaking effect}

The HQSS works well in the heavy quark limit  $m_Q\to\infty$. For finite heavy quark mass, there should be some HQSS breaking effect, which increase with the decreasing heavy quark mass. The HQSS breaking effect can be estimated in both charm and bottom sectors in a comparison between the $Z_c$ and $Z_b$ systems. In the following, we analyze the HQS breaking effect from several aspects, i.e., the fit quality, the HQS breaking parameters, the wave functions, and the effective couplings. 

\subsubsection{the fit quality}
The HQS breaking in the $Z_c$ and $Z_b$ systems can be analyzed from both qualitative and quantitative perspectives. We begin with a qualitative analysis, i.e., comparing the reduced chi-square for each fit scheme.  
As shown in Fig.~\ref{line_shape_Zc},
we find that the HQS scheme cannot describe the experimental data, neither the elastic channels nor the inelastic channels, 
in the $Z_c$ system, exhibiting a relatively poor fit with reduced chi-square $\chi^2/\mathrm{d.o.f.}=1.92$. When the HQS symmetry breaking parameters are switched on for both elastic and inelastic channels, the fit can be improved significantly with reduced chi-square $\chi^2/\mathrm{d.o.f.}=1.12$. This clearly suggests that HQSS breaking plays a non-negligible role in the $Z_c$ system.  

In contrast, for the $Z_b$ system, the experimental data can be almost equally well described in both HQS and HQSV schemes, as demonstrated by the fitted line shapes in Fig.~\ref{line_shape_Zb}. At the same time, the corresponding reduced chi-square $\chi^2/\mathrm{d.o.f.}$ values show little difference, which indicates that the HQSS breaking effect in the $Z_b$ system is small and can be essentially neglected.

\subsubsection{the HQS breaking parameters}

The HQS breaking effect can also be reflected in the HQS breaking parameters in a comparison with the HQS conserved ones. 
\begin{figure}
    \centering
    \includegraphics[width=0.75\linewidth]{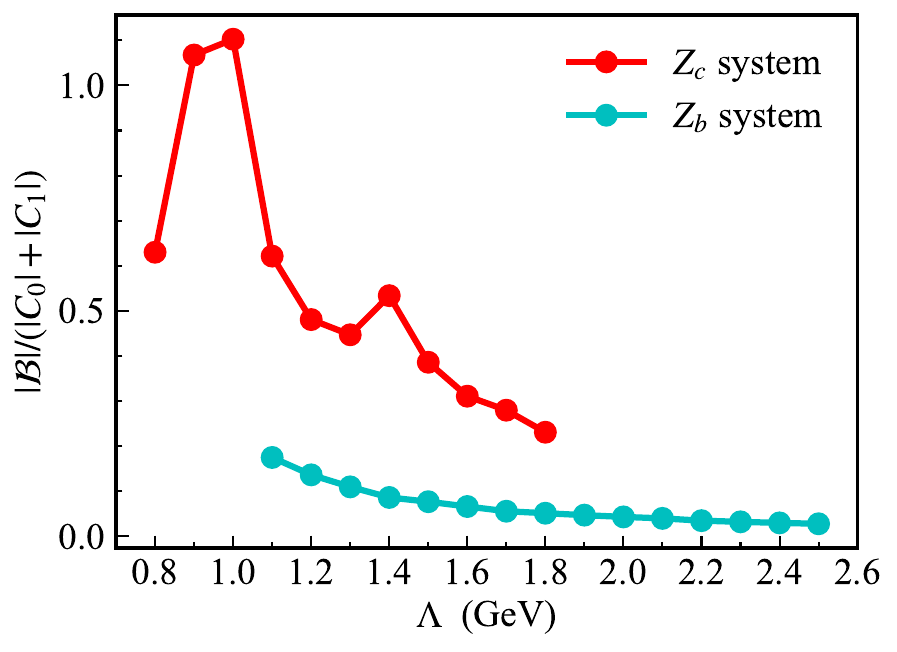}
    \caption{The cutoff dependence of  $|\mathcal{B}|/(|C_0|+|C_1|)$ for the $Z_c$ and $Z_b$ system.}
    \label{cb_rat}
\end{figure}
Fig.~\ref{cb_rat} shows the dependence of their ratio on the cutoff. The figure reveals an enhanced convergence with HQSS at larger energy scales. For the discussion in this subsection, we adopt the best-fit parameter values that yield the minimum reduced chi-square, as listed in Tab.~\ref{para}. We collect the ratios between the HQSV parameters and HQS parameters in 
Tab.~\ref{ratio}. A smaller ratio implies a better preservation of HQS, where a larger ratio indicates a more significant symmetry breaking. 
\begin{table}
\renewcommand{\arraystretch}{1.5}
    \centering
    \caption{ 
    The ratios between HQSV couplings and HQS couplings.
    The first one $|B|/(|C_0|+|C_1|)$ denotes the ratio 
    in elastic channels, while the remaining terms correspond to those in inelastic channels.}
    \label{ratio}
    \begin{tabular}{p{3cm}<{\centering}p{2.5cm}<{\centering}p{2.5cm}<{\centering}}
    \hline\hline
   $\left|C/V\right|$  &  $\bm{Z_c}$ \textbf{system} & $\bm{Z_b}$ \textbf{system}  \\
    \hline
   $|B|/(|C_0|+|C_1|) $  &  1.10  & 0.05\\
   $|v_{1S}^{\psi}/g_{1S}^{\psi}| $  &  0.37  & $-$ \\
   $|v_{1P}^{\psi}/g_{1P}^{\psi}| $  &  0.28  & $-$ \\
   $|v_{1S}^{\Upsilon}/g_{1S}^{\Upsilon}| $  &  $-$  &0.18\\
   $|v_{2S}^{\Upsilon}/g_{2S}^{\Upsilon}| $  &  $-$  &0.23\\
   $|v_{1P}^{h_b}/g_{1P}^{h_b}| $  &  $-$  & 0.20\\
   $|v_{2P}^{h_b}/g_{2P}^{h_b}| $  &  $-$  & 0.12\\
   \hline
   \hline
   \end{tabular}  
\end{table}
As shown in Tab.~\ref{ratio}, these ratios are systematically larger in the $Z_c$ system than those in the $Z_b$ system, both for the elastic and inelastic channels, indicating stronger HQSS breaking in the $Z_c$ case, which is consistent with the qualitative conclusion drawn from the analysis of the fitted line shapes.

One can see that the ratio of elastic channels in the $Z_c$ system $|B|/(|C_0|+|C_1|)>1$, which indicates a large HQS breaking effect and the reason why the HQS scheme does not work for the $Z_c$ system. 
This conclusion, that the fit quality for the HQS case in the $Z_c$ system is worse than that in the $Z_b$ system, can also be seen from 
Fig.~\ref{line_shape_Zc} and Fig.~\ref{line_shape_Zb}. This qualitative difference also suggests that the 
HQFS is less
broken in the $Z_b$ system than in the $Z_c$ system.

\subsubsection{the wave functions}
As discussed in the introduction section, the two $Z_b$ states exhibit the similar peak structures in both the  $\Upsilon(nS)\pi$ with $n=1,2,3$ and the $h_b(mP)$ with $m=1,2$ invariant mass distributions. This can be easily seen from their heavy-light decompositions 
\begin{align}
    |Z_b(10610)\rangle & = -\frac{1}{\sqrt{2}} |0\otimes 1\rangle-\frac{1}{\sqrt{2}}|1\otimes0\rangle,\notag\\
    |Z_b(10650)\rangle & = \frac{1}{\sqrt{2}} |0\otimes 1\rangle-\frac{1}{\sqrt{2}}|1\otimes0\rangle.
    \label{eq:hld1}
\end{align}
As the transition between the $B\bar{B}^*$ and $B^*\bar{B}^*$ channels is small, i.e., the values of $C_0$ and $C_1$ are almost the same as shown in Tab.~\ref{para}, the $Z_b(10610)$ couples to the $\Upsilon\pi$ and $h_b\pi$ channels via the $|1\otimes 0\rangle$ and $|0\otimes 1\rangle$ components, respectively, with equal strength. That is the reason why the two $Z_b$ states behave similarly in both $\Upsilon\pi$ and $h_b\pi$ channels. These wave functions are also related to the potentials with the definition of the low-energy constants $C_0$, $C_1$ and $B$. However, from the experimental side, the two $Z_c$ states do not behave similarly in the $J/\psi\pi$ and $h_c\pi$ channels, i.e., the $Z_c(3900)$ and $Z_c(4020)$ present in the $J/\psi\pi$ and $h_c\pi$ channels, respectively. This experimental fact is attributed to the breaking of HQSS. A detailed analysis follows.

One can solve this problem by adopting an alternative parametrization scheme to further estimate the HQSS violation in the $Z_c$ system. In our framework, the HQSS breaking effects have been reflected by the HQSS breaking parameters defined in Eq.~\eqref{eb} and Eq.~\eqref{ieb}. We can also assume that the HQSS breaking is absorbed into the coefficients in front of the heavy-light basis, e.g. 
\begin{align}
    |Z_c(3900)\rangle & = a_c |0\otimes 1\rangle+b_c|1\otimes0\rangle,\notag\\
    |Z_c(4020)\rangle & = c_c |0\otimes 1\rangle+d_c|1\otimes0\rangle.
    \label{hld2}
\end{align}
without introducing additional HQSS breaking terms. The corresponding contact potential for the elastic channels can be derived from Eq.~\eqref{HQSC} and Eq.~\eqref{hld2}
\begin{align}
    V=\left(\begin{array}{cc}
       a_c^2C_{c~0}'+b_c^2C_{c~1}'  & a_cc_cC_{c~0}'+b_cd_cC_{c~1}'  \\
       a_cc_cC_{c~0}'+b_cd_cC_{c~1}'  & c_c^2C_{c~0}'+d_c^2C_{c~1}'
    \end{array}\right),
    \label{re_ct}
\end{align}
where $a_c$, $b_c$, $c_c$, and $d_c$ denote the coefficients that incorporate the HQSS breaking effects in the elastic channels and satisfy the orthonormality condition, while $C_{c~0}'$ and $C_{c~1}'$ represent the HQSS conserving coefficients with the breaking contributions absorbed into the coefficients $a_c$, $b_c$, $c_c$, and $d_c$. By substituting the fitted parameters listed in Tab.~\ref{para} into Eq.~(\ref{vct}) and combining with Eq.~(\ref{re_ct}), one can obtain
\begin{align}
    &a_c=-0.977,\quad b_c=-0.211,\quad c_c=0.211,\notag\\
    &d_c=-0.977,\quad C_{c~0}'=-13.322,\quad C_{c~1}'=18.852.
    \label{c_break}
\end{align}
If HQSS is exactly conserved, the absolute values of $a$, $b$, $c$ and $d$ are all equal to $1/\sqrt{2}$. Therefore, the larger the deviations of $|a|$, $|b|$, $|c|$ and $|d|$ from $1/\sqrt{2}$ are, the stronger the HQSS breaking is. As 
seen from Eq.~(\ref{c_break}), the magnitude of HQSS violation in the $Z_c$ system is rather pronounced. 

In our theoretical framework for the $Z_c$ system, the $J/\psi\pi$ channel couples exclusively to the heavy-light basis state $|1\otimes 0\rangle$, while the $h_c\pi$ channel couples only to $|0\otimes 1\rangle$. Given that the coefficients $|a_c|$ and $|d_c|$ are close to unity, whereas $|b_c|$ and $|c_c|$ are approximately 0.2, the $J/\psi\pi$ channel exhibits strong coupling to the $Z_c(3900)$ and only weak coupling to the $Z_c(4020)$. Conversely, the $h_c\pi$ channel couples strongly to the $Z_c(4020)$ and very weakly to the $Z_c(3900)$. Consequently, in the $J/\psi\pi$ invariant mass spectrum, only the $Z_c(3900)$ signal is observed, while in the $h_c\pi$ spectrum, a clear signal of the $Z_c(4020)$ is expected to dominate.

\subsubsection{the effective couplings}

The wave functions discussed in the previous section are related to the parameters $C_0$, $C_1$ and $B$ in the potentials. However, the parameters $C_0$, $C_1$ have large correlation to the $J/\psi\pi$ and $h_c\pi$ inelastic channels, respectively. To estimate  the HQSS breaking effect and understand why the $Z_c(3900)$ only appears in the $J/\psi\pi$ channel, 
we extract the effective couplings of a given pole to all the inelastic channels, 
which are collected in Tab.~\ref{z_coupling}. 
From the table, one can see that the $Z_c(3900)$ couples much more strongly to the $J/\psi\pi$ channel than to the $h_c\pi$ channel. This naturally explains why it appears as a significant peak structure in the $J/\psi\pi$ channel instead of the $h_c\pi$ channel. 
However, by combining the couplings in Tab.~\ref{z_pole} and Tab.~\ref{z_coupling}, one can see that the effective coupling of the $Z_c(3900)$ to the elastic channel is significantly stronger than that to the inelastic channels, suggesting that the elastic channels are the dominant ones.

\begin{table}
\renewcommand{\arraystretch}{1.5}
    \centering
    \caption{Effective couplings of the poles (in units of GeV) to the inelastic channels in the $Z_c$ and $Z_b$ systems. The coupling constants have the dimension of $[\mathrm{GeV}]^{-1}$. The labels V and S correspond to the HQSV and HQS  
   schemes, respectively.}
    \label{z_coupling}
    \begin{tabular}{p{2.8cm}<{\centering}|p{1.2cm}<{\centering}p{1.2cm}<{\centering}p{1.2cm}<{\centering}p{1.2cm}<{\centering}}
    \hline\hline
    \textbf{Pole}$(Z_c)$&  \multicolumn{2}{c}{$g_{J/\psi\pi}$} &\multicolumn{2}{c}{$g_{h_c\pi}$}\\
    \hline
    $3.875-0.011i$~(V)&  \multicolumn{2}{c}{0.19} &\multicolumn{2}{c}{0.06}\\
    \hline
   \textbf{Pole}$(Z_b)$&  $g_{\Upsilon(1S)\pi}$ & $g_{\Upsilon(2S)\pi}$ &$g_{h_b(1P)\pi}$ & $g_{h_b(2P)\pi}$\\
    \hline
   $10.610-0.010i$~(V)  &  0.17  & 0.02 & 0.02 & 0.55\\
   $10.640-0.021i$~(V)  &  0.26  & 0.03 & 0.01 & 0.22\\
   \hline
   $10.600-0.013i$~(S)  &  0.05  & 0.19 & 0.02 &0.50 \\
   $10.649-0.016i$~(S)  &  0.06  & 0.24 & 0.02 &0.37 \\
   \hline
   \hline
   \end{tabular}  
\end{table}

\begin{figure*}
    \centering
    \includegraphics[width=0.35\linewidth]{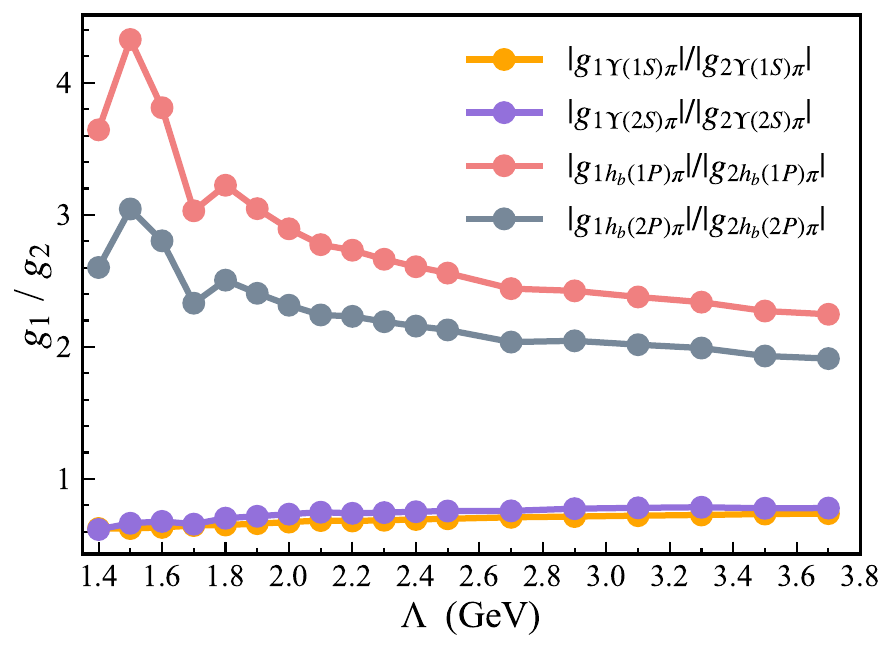}
    \includegraphics[width=0.35\linewidth]{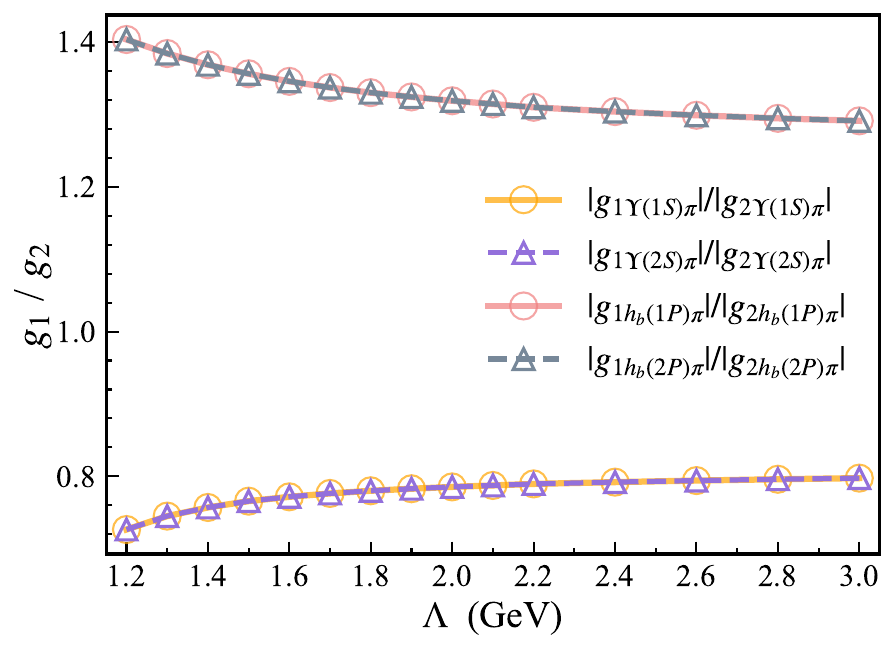}
    \caption{ 
    Ratios of the $Z_b(10610)$ effective coupling constants to  the $Z_b(10650)$ effective coupling constants at different cutoff values.
    Here, $g_1$ and $g_2$ denote the effective couplings of the $Z_b(10610)$ and $Z_b(10650)$ to the inelastic channels, respectively. The left panel shows the variation of $g_1/g_2$ with the energy scale in the HQSV 
    scheme, where the scale ranges from 1.4 to 3.7 GeV with a step of 0.1 GeV below 2.5 GeV and 0.2 GeV above. The right panel presents the HQS scheme, with the scale varying from 1.2 to 3.0 GeV, and steps of 0.1 GeV below 2.2 GeV and 0.2 GeV above.}
    \label{ratio_g}
\end{figure*}

For the $Z_b$ case,  Tab.~\ref{z_coupling} shows that, 
in the HQS fitting scheme, the effective couplings of $Z_b(10610)$ and $Z_b(10650)$ to both the $\Upsilon(nS)\pi$ and $h_b(mP)\pi$ channels are  nearly comparable, which is a natural consequence of HQSS. In the HQSV fitting scheme, the effective couplings of both  $Z_b(10610)$ and $Z_b(10650)$ to the $\Upsilon(nS)\pi$ channels are of comparable magnitude. In contrast, the effective couplings of the $Z_b(10610)$ to the $h_b(mP)\pi$ channels are roughly two to three times larger than 
those of the $Z_b(10650)$.
At first glance, the large difference between the effective couplings of $Z_b(10610)$ and $Z_b(10650)$ to the $h_b(mP)\pi$ channels in the HQSV fit seems to imply a noticeable HQSS breaking in the $Z_b$ system. However, the practical degree of HQSS violation is not as significant as it appears. This conclusion is supported by the following consideration:
\begin{enumerate}
    \item  In the $Z_b$ system, a total of five HQSS breaking parameters are introduced for the elastic and inelastic channels. The fitting results obtained with and without these five parameters are nearly identical, indicating that these HQSS breaking terms are likely redundant.
    \item As shown in Table~\ref{para}, the uncertainties of these five HQSS breaking parameters are relatively large. In particular, the uncertainty of the breaking term in the $h_b\pi\pi$ channel exceeds 50$\%$, further implying the redundancy of these parameters.
\end{enumerate}

Therefore, we conclude that the HQSS breaking effects in the $Z_b$ system are rather small and can be 
safely neglected. Consequently, the comparable ratios of the effective couplings of $Z_b(10610)$ and $Z_b(10650)$ to the $\Upsilon(nS)\pi$ and $h_b(mP)\pi$ channels naturally lead to clear signals of both states in the corresponding invariant mass distributions. Furthermore, one can also obtain the same conclusion from Fig.~\ref{ratio_g}, which shows how the ratios of the effective couplings of $Z_b(10610)$ and $Z_b(10650)$ to the $\Upsilon(nS)\pi$ and $h_b(mP)\pi$ channels vary with the cutoff. The right panel of Fig.~\ref{ratio_g} shows that the ratios go to a constant with the increasing cutoff. That ratio in the $\Upsilon(1S)\pi$ ($h_b(1P)\pi$) channel is  identical to that in the $\Upsilon(2S)\pi$ ($h_b(2P)\pi$). This behavior  exactly reflects HQSS. For the HQSV fitting scheme, the ratios also tend to constants with the increasing cutoffs. The ratio in the $\Upsilon(1S)\pi$ ($h_b(1P)\pi$) channel is not identical to that in the $\Upsilon(2S)\pi$ ($h_b(2P)\pi$) anymore. The fluctuations at small cutoffs stem from the correlation between the inelastic channels and the HQSS breaking parameters.

%% file: Summary.tex
\section{SUMMARY}\label{Summary}
In this work, we perform a phenomenological study of the invariant mass spectra in both the $Z_c$ and $Z_b$ systems, including the elastic and inelastic channels. Within the framework of HQSS and its breaking effects, we construct the $S$-wave interactions between the elastic channels as well as between the elastic and inelastic channels. Combined with the LSE, two distinct schemes are employed to perform a global fit to the above invariant mass spectra. 

From the fit, we extract the pole positions of the $Z_c$ and $Z_b$ systems under both HQS and HQSV 
schemes. To examine model dependence, the Gaussian form factor in the two-point function is replaced with a monopole form, 
but both cases give rise to consistent line shapes, reduced chi-squares, and pole positions—indicating weak model sensitivity. The stability of the results is further confirmed by refits at different cutoffs, which show only minor variations in the pole locations. In particular, for the $Z_b$ system in HQS 
scheme, both the real and imaginary parts of the poles converge as the cutoff increases, demonstrating the robustness of our framework.

We then evaluate the effective couplings of these poles to the inelastic channels. For the $Z_b$ system, HQSS breaking effects are negligible, and the effective coupling strengths of $Z_b(10610)$ and $Z_b(10650)$ to the inelastic channels are comparable. Consequently, their similar coupling ratios 
in the $\Upsilon(nS)\pi$ and $h_b(mP)\pi$ channels naturally explain the clear signals of both states in the corresponding invariant-mass distributions. Moreover, the cutoff dependence of these ratios shows that HQSS becomes increasingly well satisfied as the cutoff grows.

For the $Z_c$ system, we calculate the effective couplings of $Z_c(3900)$ to the $J/\psi\pi$ and $h_c\pi$ channels in the HQSV case. Although the absence of the $Z_c(4020)$ pole prevents a direct comparison, the analysis indicates that HQSS breaking mainly occurs in the elastic channels. 
The effective couplings of $Z_c(3900)$ to $J/\psi\pi$ and $h_c\pi$ indicates that 
it couples much more strongly to $J/\psi\pi$. Therefore, only the $Z_c(3900)$ signal is expected in the $J/\psi\pi$ spectrum.

In summary, this work provides a coherent dynamical description of the $Z_c$ and $Z_b$ families, establishing the HQSS breaking as the key to understanding their distinct behaviors. The $Z_b$ states are well-defined molecular states under HQSS, while $Z_c(3900)$ is a molecular state accompanied by significant symmetry breaking, and $Z_c(4020)$  may not be a conventional resonance. The robustness of the model has been verified through variations in the form factor and cutoff, reinforcing the reliability of the conclusions.

%% file: supp.tex
\section{Three-body decay}\label{three_body}
\begin{figure}[h]
    \centering 
    \subfigure[]{
    \includegraphics[width=0.35\textwidth]{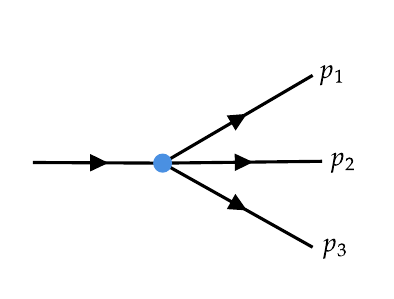}
    \label{three_1}
    }
    \subfigure[]{
    \includegraphics[width=0.35\textwidth]{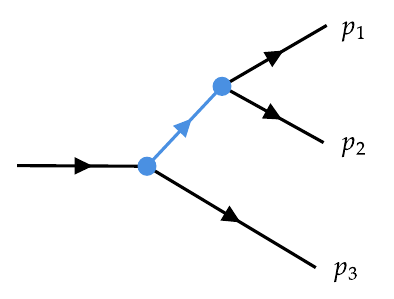}
    \label{three_2}
    }
    \caption{Schematic diagram of the three-body decay. (a) The process of $X \to a + b + c$. (b) The process of $X \to Y +c \to a + b + c$, where $Y$ is an intermediate state.}
    \label{three}
\end{figure}
In this work, we do not take into account the three-body final-state interactions. Instead, we restrict our analysis to the quasi-two-body mechanism illustrated in Fig.~\ref{three_2}, where the parent particle first couples to an intermediate two-body subsystem, followed by the emission of the third particle. This treatment significantly simplifies the theoretical framework while still capturing the dominant kinematic features of the process. 

To describe the kinematics of the decay process, we introduce the invariant mass variables of the two-body subsystems, defined as 
\begin{align}
    p_{ij}=p_i+p_j,\quad s_{ij}=m_{ij}^2=p_{ij}^2,
\end{align}
where $p_i$ is the four-momentum of the $i$-th final-state particle. In the rest frame of particles 1 and 2, the energies and  the three momenta are given by
\begin{align}
    &E_1^*  =\frac{m_{12}^2+m_1^2-m_2^2}{2m_{12}},\quad \quad E_2^* =\frac{m_{12}^2+m_2^2-m_1^2}{2m_{12}},\quad \quad E_3^* =\frac{M^2-m_{12}^2-m_3^2}{2m_{12}},\notag\\
    &|p_1^*|  = |p_2^*|  = \frac{\lambda^{1/2}(m_{12}^2,m_1^2,m_2^2)}{2m_{12}},\quad \quad |p_3^*|=\frac{\lambda^{1/2}(M^2,m_{12}^2,m_3^2)}{2m_{12}},
\end{align}
where $\lambda$ is the K\"all\'en function. Similarly, in the rest frame of the parent particle with mass $M$, the energies and momenta of the final-state particles are expressed as
\begin{align}
    E_1 & =\frac{M^2+m_1^2-m_{23}^2}{2M},\quad \quad E_2 =\frac{M^2+m_2^2-m_{13}^2}{2M},\quad \quad E_3 =\frac{M^2+m_3^2-m_{12}^2}{2M},\notag\\
    |p_1| & =\frac{\lambda^{1/2}(M^2,m_1^2,m_{23}^2)}{2M},\quad \quad |p_2|  =\frac{\lambda^{1/2}(M^2,m_2^2,m_{13}^2)}{2M},\quad \quad |p_3|  =\frac{\lambda^{1/2}(M^2,m_3^2,m_{12}^2)}{2M}.
\end{align}
The three invariant masses $m_{12}^2$, $m_{23}^2$ and $m_{13}^2$ are not independent but satisfy the kinematic constraint
\begin{align}
m_{12}^2+m_{13}^2+m_{23}^2=M^2+m_1^2+m_2^2+m_3^2,
\end{align}
which implies that only two of them can be chosen as independent variables. The physical range for each invariant mass is determined by
\begin{align}
    m_{ij}^2 \in [(m_i+m_j)^2,(M-m_k)^2],\quad (i\neq j \neq k).
\end{align}

The general expression of the $n$-body phase space is
\begin{align}
    d\Phi_{n} = \prod_{i=1}^{n} \frac{d^3p_i}{(2\pi)^3 2E_i}\delta^4\left(p_a+p_b-\sum_i p_i\right),
\end{align}
where $p_a$ and $p_b$ are the four momenta of the parent particles, and $p_i$ is the four momentum of the final state. For the three-body case, this expression can be rewritten as
\begin{align}
    d\Phi_3 = & \frac{d^3p_1}{(2\pi)^3 2E_1}\frac{d^3p_2}{(2\pi)^3 2E_2}\frac{d^3p_3}{(2\pi)^3 2E_3}(2\pi)^4\delta^4(P-p_1-p_2-p_3)\notag\\
    = & \frac{1}{4(2\pi)^3}dE_1dE_2 = \frac{1}{16(2\pi)^3}\frac{dm_{23}^2dm_{13}^2}{M^2} = \frac{1}{16(2\pi)^3}\frac{dm_{12}^2dm_{13}^2}{M^2} =\frac{1}{16(2\pi)^3}\frac{dm_{12}^2dm_{23}^2}{M^2}
\end{align}
When the reaction is dominated by a quasi-two-body mechanism, the intermediate state formed by particles 1 and 2 plays the leading role in the dynamics. In this situation, the invariant mass $m_{12}$ can be regarded as an independent kinematic variable, while the remaining degrees of freedom, such as $m_{23}$, only contribute through the available phase space. Consequently, the differential decay width with respect to $m_{12}$ is written as
\begin{align}
    \frac{d\Gamma}{dm_{12}} = \frac{1}{16(2\pi)^3M^3}\int_{(m_{23}^2)_{min}}^{(m_{23}^2)_{max}}m_{12}|\mathcal{M}(m_{12})|^2dm_{23}^2,
    \label{mass_dis}
\end{align}
where $\mathcal{M}(m_{12})$ denotes the decay amplitude, which in this approximation only depends on the two-body subsystem of particles 1 and 2. The integration limits of $m_{23}^2$ are determined by kinematic constraints in the rest frame of the $1$–$2$ system and can be written as follow
\begin{align}
    (m_{23}^2)_{max} & = (E_2^*+E_3^*)^2-(\sqrt{E_2^{*2}-m_2^2}-\sqrt{E_3^{*2}-m_3^2})^2\notag\\
    & = (E_2^*+E_3^*)^2 - \frac{1}{4m_{12}^2}[\lambda^{1/2}(m_{12}^2,m_1^2,m_2^2)-\lambda^{1/2}(M^2,m_{12}^2,m_3^2)]^2,\notag\\
    (m_{23}^2)_{min} & = (E_2^*+E_3^*)^2-(\sqrt{E_2^{*2}-m_2^2}+\sqrt{E_3^{*2}-m_3^2})^2\notag\\
    & = (E_2^*+E_3^*)^2 - \frac{1}{4m_{12}^2}[\lambda^{1/2}(m_{12}^2,m_1^2,m_2^2)+\lambda^{1/2}(M^2,m_{12}^2,m_3^2)]^2.
    \label{limits}
\end{align}
Substituting Eq.~\eqref{limits} into Eq.~\eqref{mass_dis} and carrying out the integration over $m_{23}^2$, one can obtain
\begin{align}
    \frac{d\Gamma}{dm_{12}} & = \frac{1}{16(2\pi)^3M^3}\int_{(m_{23}^2)_{min}}^{(m_{23}^2)_{max}}m_{12}|\mathcal{M}(m_{12})|^2dm_{23}^2\notag\\
    & =\frac{1}{16(2\pi)^3M^3}m_{12}|\mathcal{M}(m_{12})|^2\left[(m_{23}^2)_{max}-(m_{23}^2)_{min}\right]\notag\\
    & = \frac{1}{16(2\pi)^3M^3}|\mathcal{M}(m_{12})|^2\frac{\lambda^{1/2}(m_{12}^2,m_1^2,m_2^2)\lambda^{1/2}(M^2,m_{12}^2,m_3^2)}{m_{12}}\notag\\
    & = \frac{p_1^*p_3}{32\pi^2M^2}|\mathcal{M}(m_{12})^2|,
\end{align}
where $p_1^*$ is the momentum of particle 1 in the $1$–$2$ rest frame and $p_3$ is the momentum of particle~3 in the parent rest frame. More details can be found in 
\cite{ParticleDataGroup:2022pth, Niu:2024cfn}.

\newpage
\section{Fitted parameters for $Z_c$ and $Z_b$ system}\label{FIT}
\begin{table*}[hbt!]
\renewcommand{\arraystretch}{1.5}
    \centering
    \caption{The fitted parameters for the $Z_c$ and $Z_b$ systems. The coefficients $C_0$, $C_1$, and $\mathcal{B}$ correspond to the interaction potentials 
    between the elastic channels, while the parameters $g$ and $v$ describe the interactions 
    between the elastic channels and the inelastic channels. The parameters $U^{\text{c~(b)}}$ represent the bare production amplitude at $\sqrt{s}=4.26$~GeV for the $Z_c$ system (or the bare production amplitude for the $Z_b$ system), while $U'^{\text{c}}$ correspond to the bare production amplitude of the $Z_c$ system at $\sqrt{s}=4.23$~GeV. The factors $\mathcal{N}$ are normalization coefficients for the elastic and inelastic channels, and $\Lambda$ denotes the energy scale of the system.}
    \label{para_A}
    \begin{tabular}{p{3.3cm}<{\centering}|p{3.3cm}<{\centering}p{3.3cm}<{\centering}|p{3.3cm}<{\centering}p{3.3cm}<{\centering}}
    \hline\hline
   \textbf{Parameters}  &  \textbf{HQSV}~$(Z_c)$ & \textbf{HQS}~$(Z_c)$ & \textbf{HQSV}~$(Z_b)$ & \textbf{HQS}~$(Z_b)$  \\
    \hline
   $C_0~[\mathrm{GeV^{-2}}] $  &  $-3.88\pm 0.37$  &$-6.75\pm 0.13$& $-2.21\pm 0.06$ & $-2.64\pm 0.12$  \\
   $C_1~[\mathrm{GeV^{-2}}]$  &  $9.41\pm 1.00$ &$-3.32\pm 0.29$&  $-2.67 \pm 0.07$   & $-2.88\pm 0.11$\\
   $\mathcal{B}~[\mathrm{GeV^{-2}}]$  & $-14.65\pm 0.46$ &$-$& $-0.25\pm 0.04$ &$-$  \\ 
   $g_{1S}^{\psi}~[\mathrm{GeV^{-3/2}}]$  & $-5.34\pm 0.09$ &$-2.15\pm 0.07$& $-$ & $-$\\
   $g_{1P}^{h_c}~[\mathrm{GeV^{-2}}]$  &  $0.98\pm 0.06$ &$-0.31\pm 0.04$& $-$ & $-$ \\
   $v_{1S}^{\psi}~[\mathrm{GeV^{-3/2}}]$  & $1.99\pm 0.12$ &$-$&  $-$ & $-$ \\
   $v_{1P}^{\psi}~[\mathrm{GeV^{-2}}]$  & $0.27\pm 0.02$ &$-$&  $-$ & $-$ \\
   $g_{1S}^{\Upsilon}~[\mathrm{GeV^{-3/2}}]$  & $-$ & $-$& $-1.42\pm 0.05$ & $-0.47\pm 0.12$ \\
   $g_{2S}^{\Upsilon}~[\mathrm{GeV^{-3/2}}]$  & $-$  &$-$  & $-0.13\pm 0.02$ & $-1.99\pm 0.08$ \\
   $g_{1P}^{h_b}~[\mathrm{GeV^{-2}}]$  & $-$  & $-$ & $0.15 \pm 0.04$ & $0.24\pm 0.06$ \\
   $g_{2P}^{h_b}~[\mathrm{GeV^{-2}}]$  & $-$  & $-$ & $-4.30 \pm 0.51$ & $-4.95\pm 0.64$ \\
   $v_{1S}^{\Upsilon}~[\mathrm{GeV^{-3/2}}]$  & $-$ & $-$& $-0.26\pm 0.07$ & $-$ \\
   $v_{2S}^{\Upsilon}~[\mathrm{GeV^{-3/2}}]$  & $-$  & $-$ & $-0.03\pm 0.01$ & $-$ \\
   $v_{1P}^{h_b}~[\mathrm{GeV^{-2}}]$  & $-$  & $-$ & $0.03 \pm 0.01$ &$-$  \\
   $v_{2P}^{h_b}~[\mathrm{GeV^{-2}}]$  & $-$  &$-$  & $-0.50 \pm 0.28$ & $-$ \\
   $U_{1}^\text{c~(b)}~[\mathrm{GeV^{0}}]$  & $534.45\pm 22.41$  &$85.25\pm 22.27$  & $-2256.56\pm 210.77$ & $-3289.53\pm 296.67 $ \\
   $U_{2}^\text{c~(b)}~[\mathrm{GeV^{0}}]$  & $431.11\pm 61.63$  &$303.31\pm 21.37$  & $1620.66\pm 544.53$ & $1122.08\pm 340.03$ \\
   $U_{3}^\text{c~(b)}~[\mathrm{GeV^{-1}}]$  & $-1234.81\pm 391.94$  &$-7693.26\pm 310.46$  & $-27527.88\pm 8734.01$ & $-56250.68\pm 10842.96$ \\
   $U_{4}^\text{c~(b)}~[\mathrm{GeV^{-1}}]$  & $68849.04\pm 1888.32$  &$6236.14\pm 586.43$  & $253771.04\pm 30576.39$ & $133616.87\pm 18340.85$ \\
   $U_{1}'^{\text{c}}~[\mathrm{GeV^{0}}]$  & $-739.40\pm 154.78$  &$108.58\pm 66.72$  & $-$ & $-$ \\
   $U_{2}'^{\text{c}}~[\mathrm{GeV^{0}}]$  & $1375.14\pm 156.44$  &$637.08\pm 67.28$  & $-$ & $-$ \\
   $U_{3}'^{\text{c}}~[\mathrm{GeV^{-1}}]$  & $8711.76\pm 1905.98$  &$-17895.19\pm 1064.27$  & $-$ & $-$ \\
   $U_{4}'^{\text{c}}~[\mathrm{GeV^{-1}}]$  & $98799.25\pm 3628.51$  &$-1725.35\pm 1537.01$  & $-$ & $-$ \\
   $\mathcal{N}_{D^*\bar{D}^*}~[\mathrm{GeV^{0}}]$  & $-12.71\pm 1.06$  &$-10.44\pm 0.89$  & $-$ & $-$ \\
   $\mathcal{N}_{J/\psi\pi}~[\mathrm{GeV^{0}}]$  & $27.66\pm 1.34$  &$66.10\pm 4.28$  & $-$ & $-$ \\
   $\mathcal{N}_{h_c\pi}~[\mathrm{GeV^{0}}]$  & $-61738.51\pm 10495.42$  &$-22481.63\pm 5302.66$  & $-$ & $-$ \\
   $\mathcal{N}_{B^*\bar{B}^*}~[\mathrm{GeV^{0}}]$  & $-$  &$-$  & $0.39\pm 0.09$ & $0.95\pm 0.12$ \\
   $\mathcal{N}_{\Upsilon(1S)\pi}~[\mathrm{GeV^{0}}]$  & $-$  &$-$  & $0.31\pm 0.04$ & $7.79\pm 4.14$ \\
   $\mathcal{N}_{\Upsilon(2S)\pi}~[\mathrm{GeV^{0}}]$  & $-$  &$-$  & $145.66\pm 41.04$ & $1.68\pm 0.17$ \\
   $\mathcal{N}_{h_b(1P)\pi}~[\mathrm{GeV^{0}}]$  & $-$  &$-$  & $55315.24\pm 26055.55$ & $33814.43\pm 16367.59$ \\
   $\mathcal{N}_{h_b(2b)\pi}~[\mathrm{GeV^{0}}]$  & $-$  &$-$  & $178.40\pm 56.52$ & $203.86\pm 59.41$ \\
   $\Lambda~[\mathrm{GeV^{1}}]$  & $1.00$  &$1.80$  & $1.80$ & $1.40$ \\
   $\chi^2/\mathrm{d.o.f.}$ & 1.12  & 1.92 & 1.76 & 1.81 \\
   \hline
   \hline
   \end{tabular}  
\end{table*}

\newpage
\section{The distributions of the standardized residuals for the $Z_c$ and $Z_b$ system}\label{residual}
\begin{figure}[hbt!]
    \centering
    \includegraphics[width=0.35\linewidth]{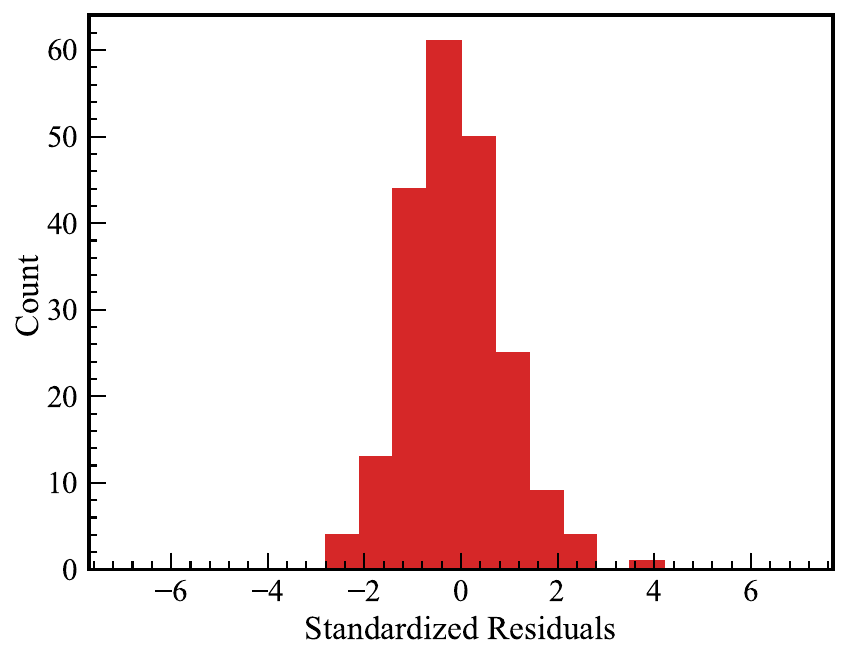}
    \includegraphics[width=0.35\linewidth]{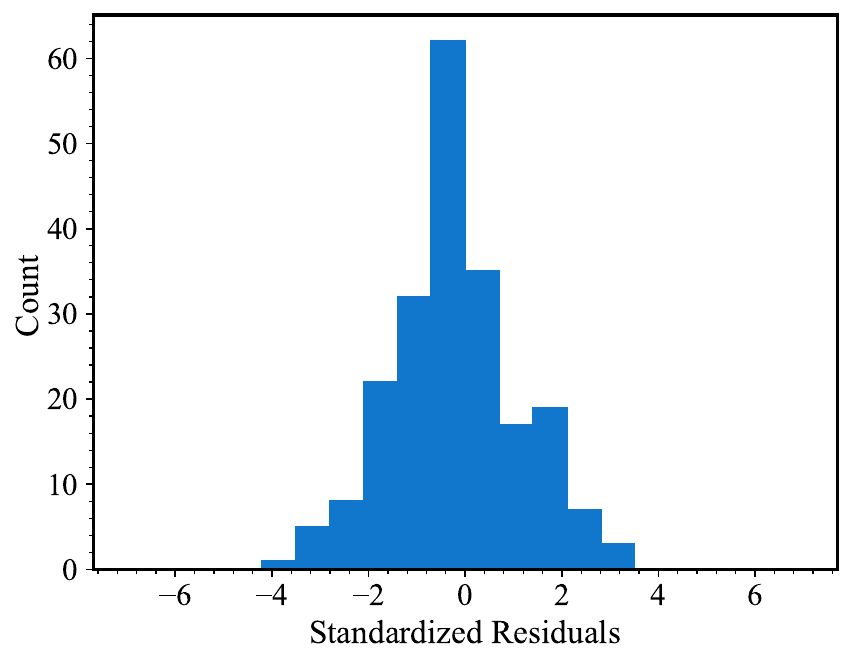}
    \caption{The distributions of the standardized residuals for the HQSV (red) and HQS (blue) schemes in the $Z_c$ system. The residuals are binned into 20 intervals within the range of [-7, 7].}  
    \label{residual_Zc}
\end{figure}

\begin{figure}[hbt!]
    \centering
    \includegraphics[width=0.35\linewidth]{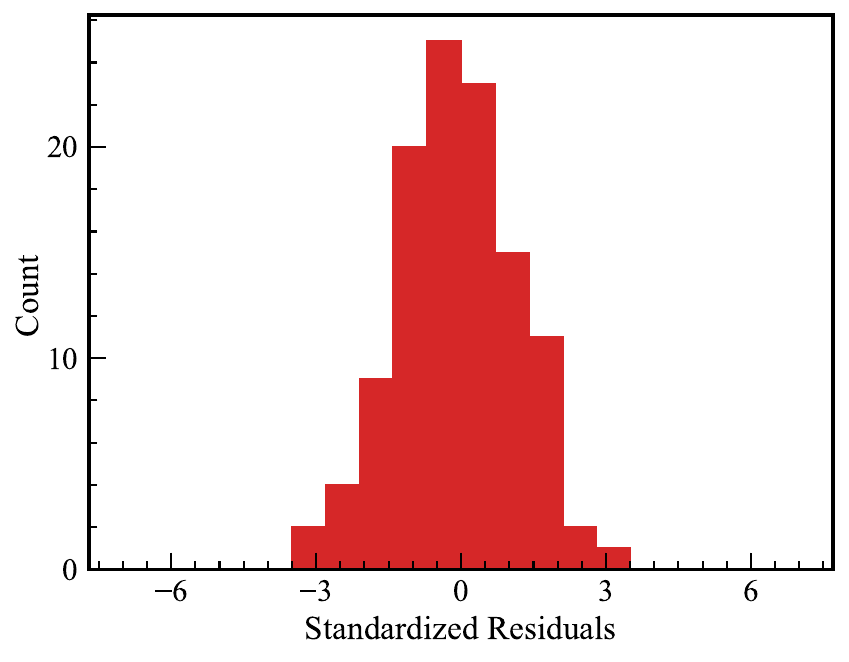}
    \includegraphics[width=0.35\linewidth]{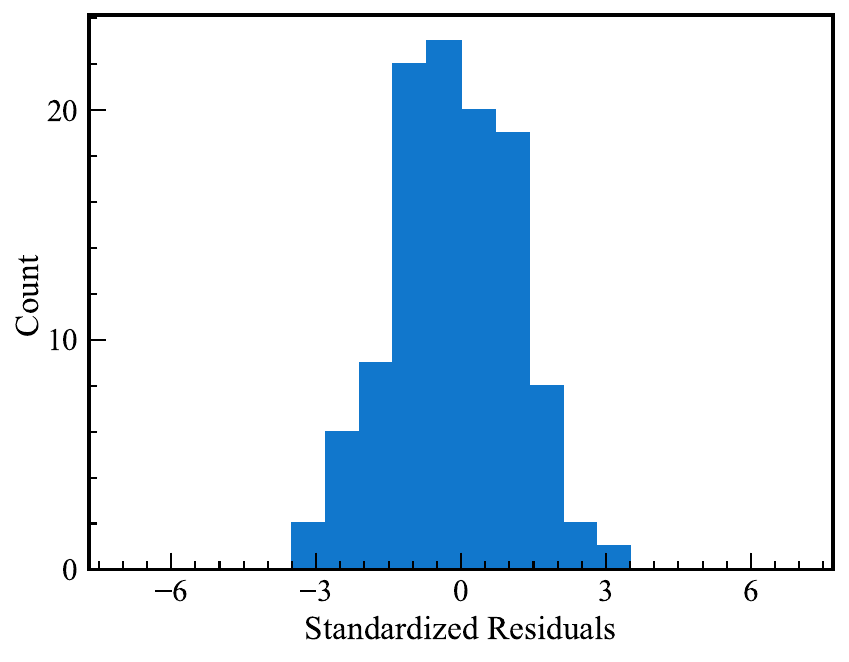}
    \caption{The distributions of the standardized residuals for the HQSV (red) and HQS (blue) schemes in the $Z_b$ system. The residuals are binned into 20 intervals within the range of [-7, 7].}   
    \label{residual_Zb}
\end{figure}
As presented in Fig.~\ref{residual_Zc} and Fig.~\ref{residual_Zb}, the HQSV fit in the $Z_c$ system and both fitting schemes in the $Z_b$ system yield residuals that are well described by a Gaussian distribution, indicating statistically acceptable fits. In contrast, the HQS fit in the $Z_c$ system shows a clear deviation from Gaussian behavior, suggesting an inferior fit quality.

\section{Examination of the model stability with a monopole form factor}\label{model_d}
To examine the stability of the model, we replace the Gaussian form factor in the two-point functions with a monopole form factor and compare the resulting fit outcomes. Specifically, we adopt the monopole form factor 
\begin{align}
    F(q)=\frac{\Lambda^2}{\Lambda^2+\vec{q}\,^2},
\end{align}
where $\bm{q}$ denotes the three-momentum in the c.m. frame. Compared to the Gaussian form factor, the monopole form factor exhibits a slower fall-off at large momenta, which generally enhances the contribution from the ultraviolet energy region. In the nonrelativistic approximation, the two-point function can then be expressed as a three-dimensional integral over the internal momentum:
\begin{align}
    G(E) = & \int\frac{d^3q}{(2\pi)^3}\frac{F(q)^2}{E_s-\frac{\vec{q}}{2\mu}+i\varepsilon^+}\notag\\
    = & \int_0^{2\pi}d\phi\int_0^{\pi}\int_0^{+\infty}dq\frac{2\mu\vec{q}\,^2\sin(\theta)\Lambda^4}{(2\pi)^3(\Lambda^2+\vec{q}\,^2)^2(2\mu E_s-\vec{q}\,^2+i\varepsilon^+)}\notag\\
    = & -\frac{\mu \Lambda^4}{\pi^2}\int_0^{+\infty}dq\frac{\vec{q}\,^2}{(\Lambda^2+\vec{q}\,^2)^2(\vec{q}\,^2-2\mu E_s-i\varepsilon^+)}.
\end{align}
Here, $E_s=E-m_1-m_2$ is the energy measured relative to the two particle threshold, and $\mu$ is the reduced mass of the two particle system. To simplify the integrand, we decompose it into several terms, separating contributions that can be directly integrated from those that contain the pole at $\vec{q}\,^2=2\mu E_s$
\begin{align}
    G(E)
    = & -\frac{\mu \Lambda^4}{\pi^2}\int_0^{+\infty}dq\frac{\vec{q}\,^2}{(\Lambda^2+\vec{q}\,^2)^2(\vec{q}\,^2-2\mu E_s-i\varepsilon^+)}\notag\\
    = & -\frac{\mu \Lambda^4}{\pi^2}\int_0^{+\infty}\frac{dq}{(\Lambda^2+\vec{q}\,^2)^2}-\frac{2\mu^2\Lambda^4E_s}{\pi^2}\int_0^{+\infty}\frac{dq}{(\Lambda^2+\vec{q}\,^2)^2(\vec{q}\,^2-2\mu E_s-i\varepsilon^+)}\notag\\
    = & -\frac{\mu \Lambda^4}{\pi^2}\int_0^{+\infty}\frac{dq}{(\Lambda^2+\vec{q}\,^2)^2} + \frac{2\mu^2 \Lambda^4 E_s}{\pi^2(\Lambda^2+2\mu E_s)^2}\int_0^{+\infty}\frac{dq}{\Lambda^2+\vec{q}\,^2}-\frac{2\mu^2 \Lambda^4 E_s}{\pi^2(\Lambda^2+2\mu E_s)^2}\int_0^{+\infty}dq\frac{\Lambda^2}{(\Lambda^2+\vec{q}\,^2)^2}\notag\\
    & + \frac{2\mu^2 \Lambda^4 E_s(2\mu E_s+2\Lambda^2)}{\pi^2(\Lambda^2+2\mu E_s)^2}\int_0^{+\infty}\frac{dq}{(\Lambda^2+\vec{q}\,^2)^2} - \frac{2\mu^2 \Lambda^4 E_s}{\pi^2(\Lambda^2+2\mu E_s)^2} \int_0^{+\infty} \frac{dq}{\vec{q}\,^2-2\mu E_s-i\varepsilon^+}.
    \label{mono}
\end{align}
The integrals appearing in the decomposition can be evaluated using the following standard formulas
\begin{align}
    \int\frac{dx}{(a^2+x^2)^2} & =\frac{1}{2a^3}(\arctan (\frac{x}{a})+\frac{ax}{x^2+a^2})+\text{C},\notag\\
    \int\frac{dx}{a^2+x^2} & =\frac{1}{a}\arctan(\frac{x}{a})+\text{C}, 
\end{align}
together with the residue theorem to handle the pole in the complex plane. Applying these techniques, Eq.~(\ref{mono}) can be simplified as 
\begin{align}
    G(E) & =-\frac{\mu\Lambda}{4\pi}+\frac{\mu^2\Lambda^3E_s}{\pi(\Lambda^2+2\mu E_s)^2}-\frac{\mu^2\Lambda^3E_s}{2\pi(\Lambda^2+2\mu E_s)^2}+\frac{\mu^2\Lambda E_s(2\mu E_s+2\Lambda^2)}{2\pi(\Lambda^2+2\mu E_s)^2}-\frac{i\mu^2\Lambda^4 E_s}{\pi(\Lambda^2+2\mu E_s)^2\sqrt{2\mu E_s}}\notag\\
    & = \frac{\mu \Lambda^3}{4\pi}\frac{\sqrt{2\mu E_s}(2\mu E_s-\Lambda^2)-4i\Lambda \mu E_s}{(\Lambda^2+2\mu E_s)^2\sqrt{2\mu E_s}}\notag\\
    & = \frac{\mu\Lambda^3}{4\pi}\frac{1}{(k+i\Lambda)^2},
    \label{G_mono}
\end{align}
where $k=\sqrt{2\mu E_s}$. 

\begin{figure*}
    \centering
    \includegraphics[width=0.28\linewidth]{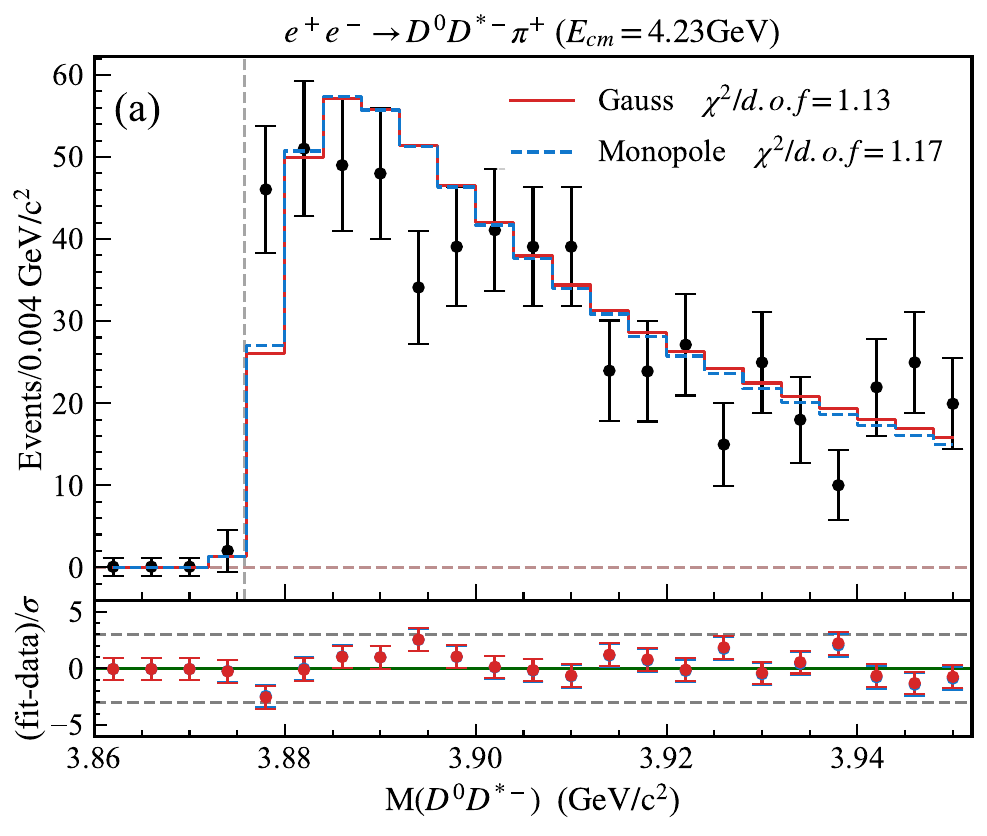}
    \includegraphics[width=0.28\linewidth]{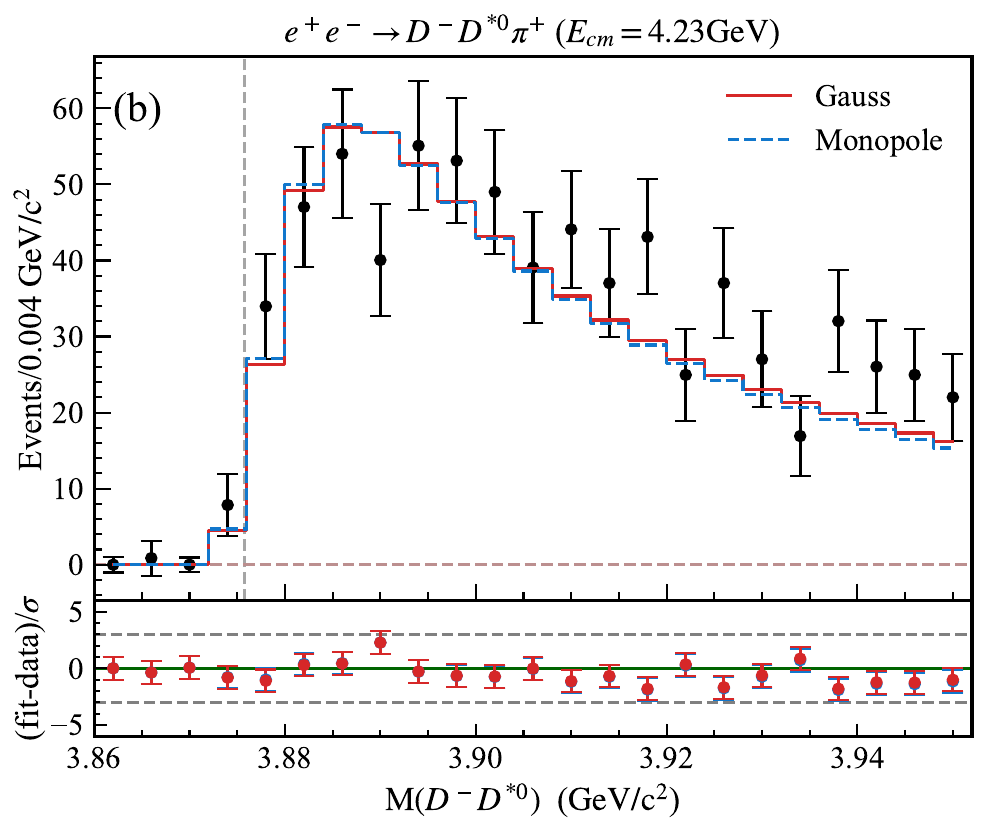}
    \includegraphics[width=0.28\linewidth]{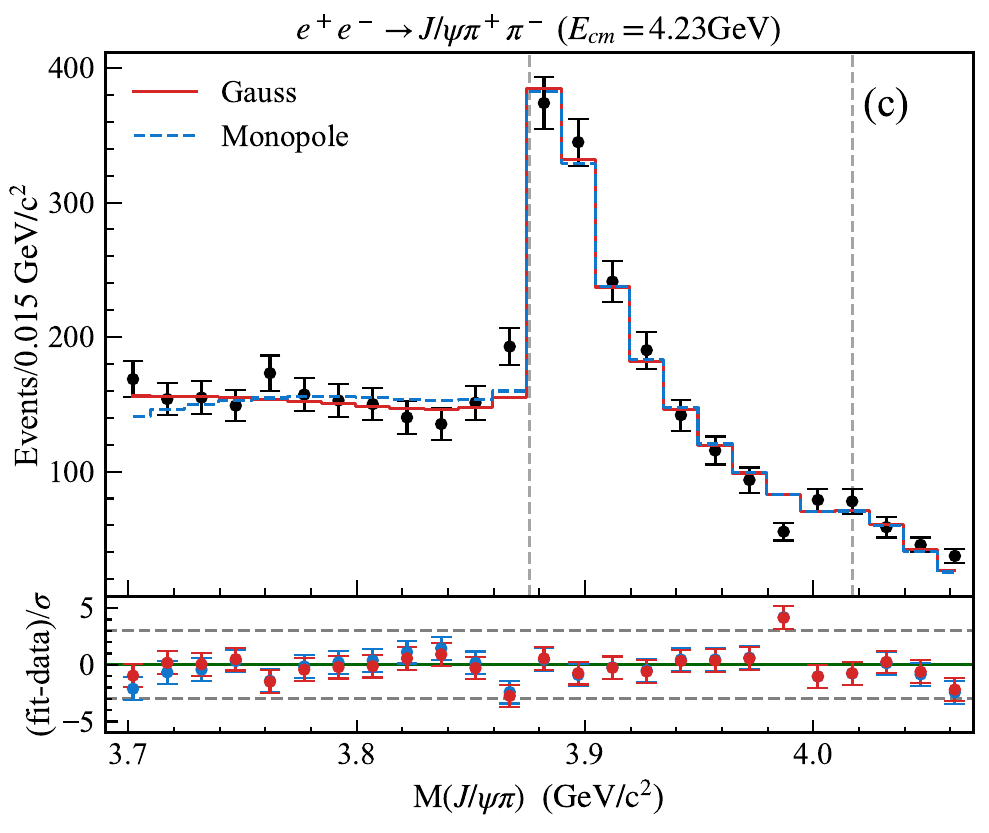}
    \includegraphics[width=0.28\linewidth]{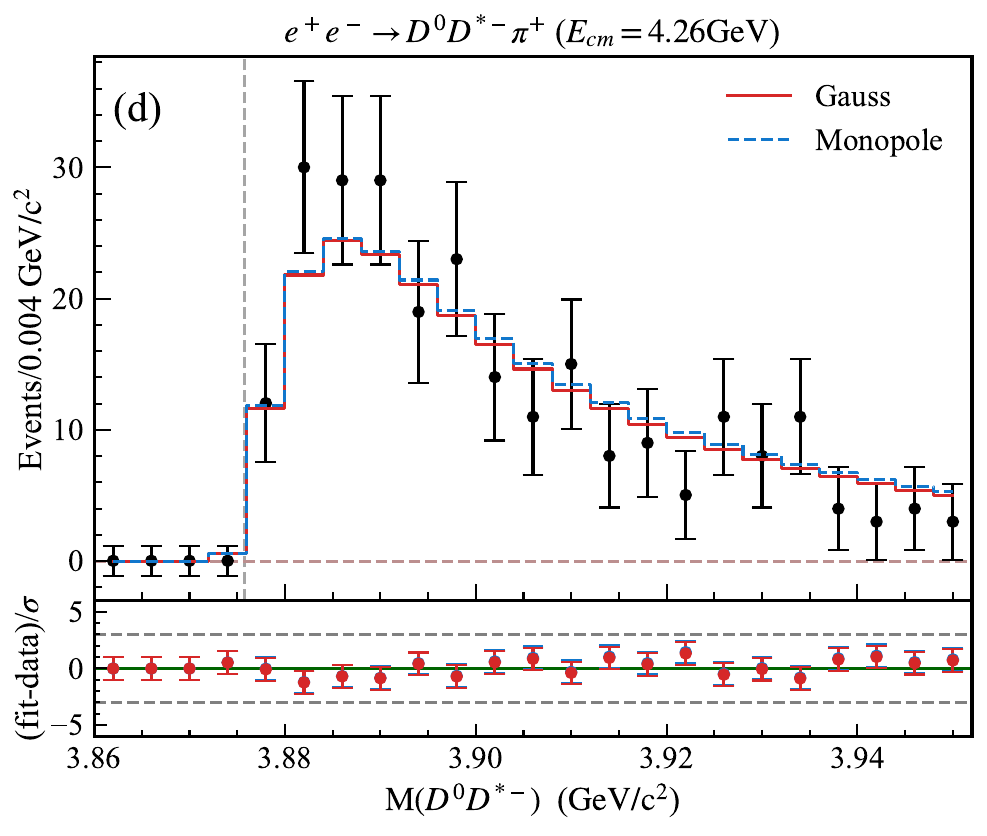}
    \includegraphics[width=0.28\linewidth]{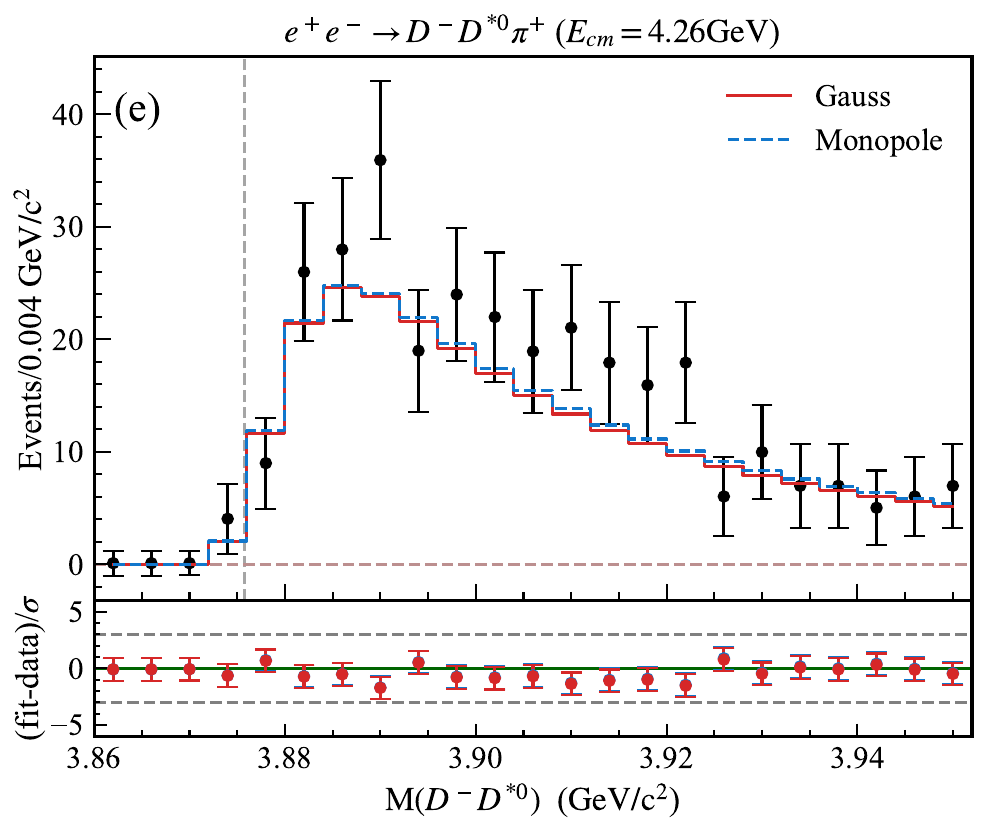}
    \includegraphics[width=0.28\linewidth]{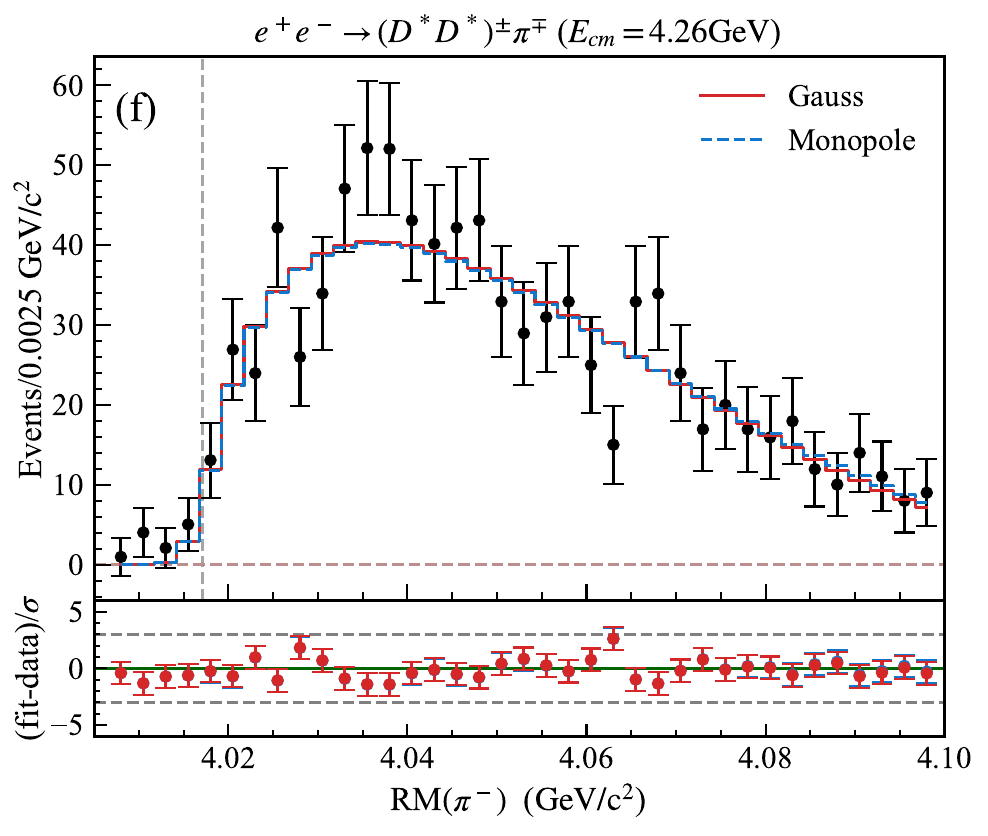}
    \includegraphics[width=0.28\linewidth]{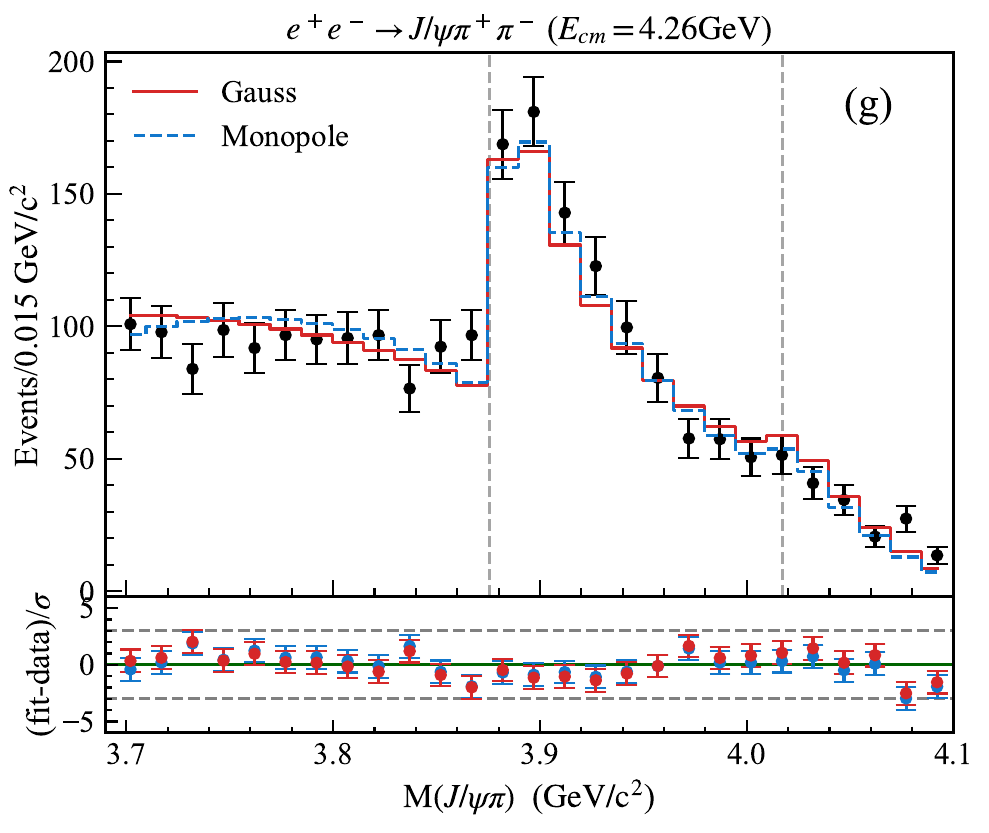}
    \includegraphics[width=0.28\linewidth]{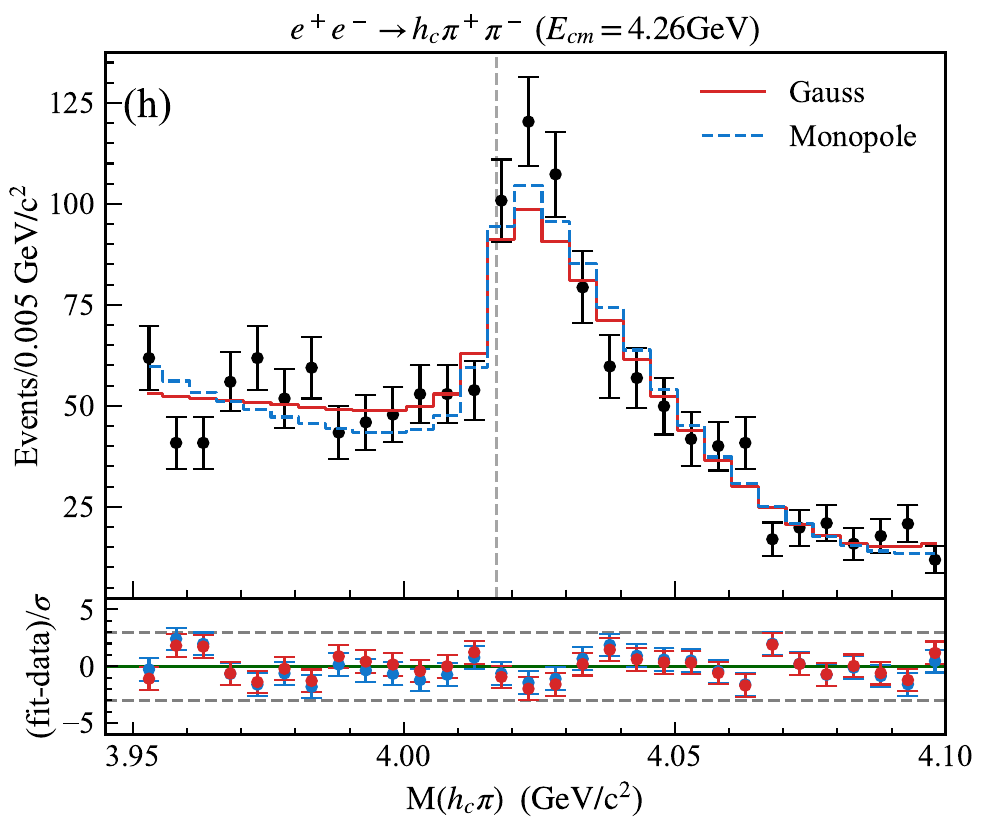}
    \includegraphics[width=0.28\linewidth]{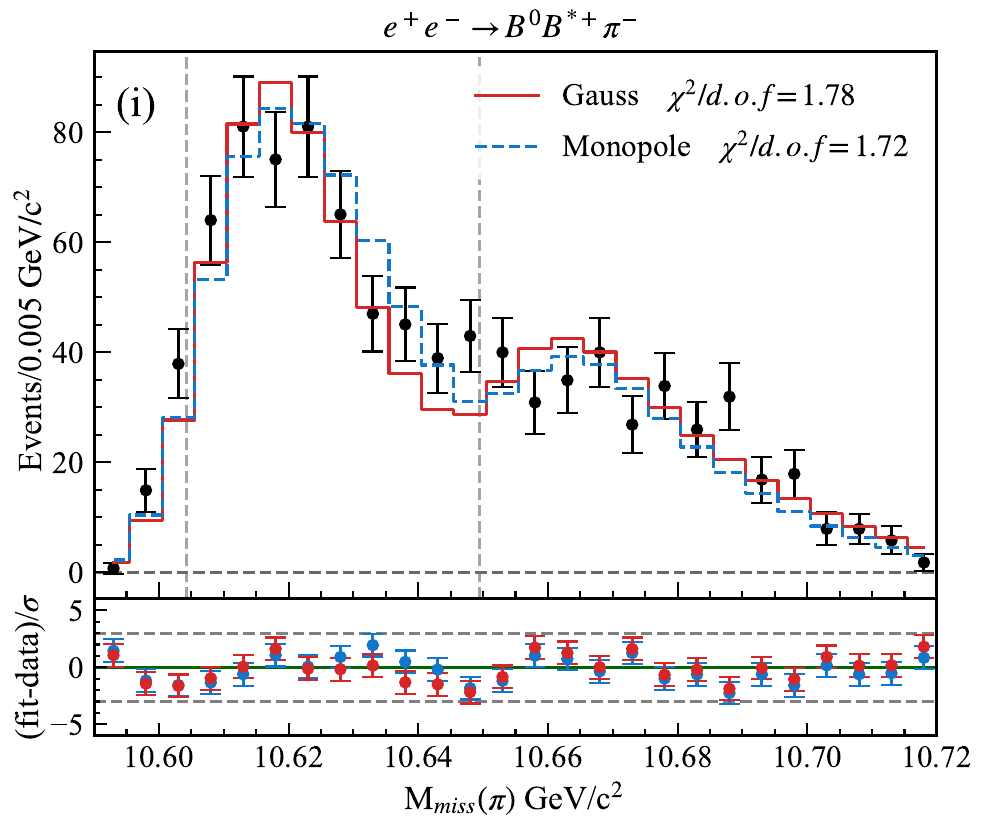}
    \includegraphics[width=0.28\linewidth]{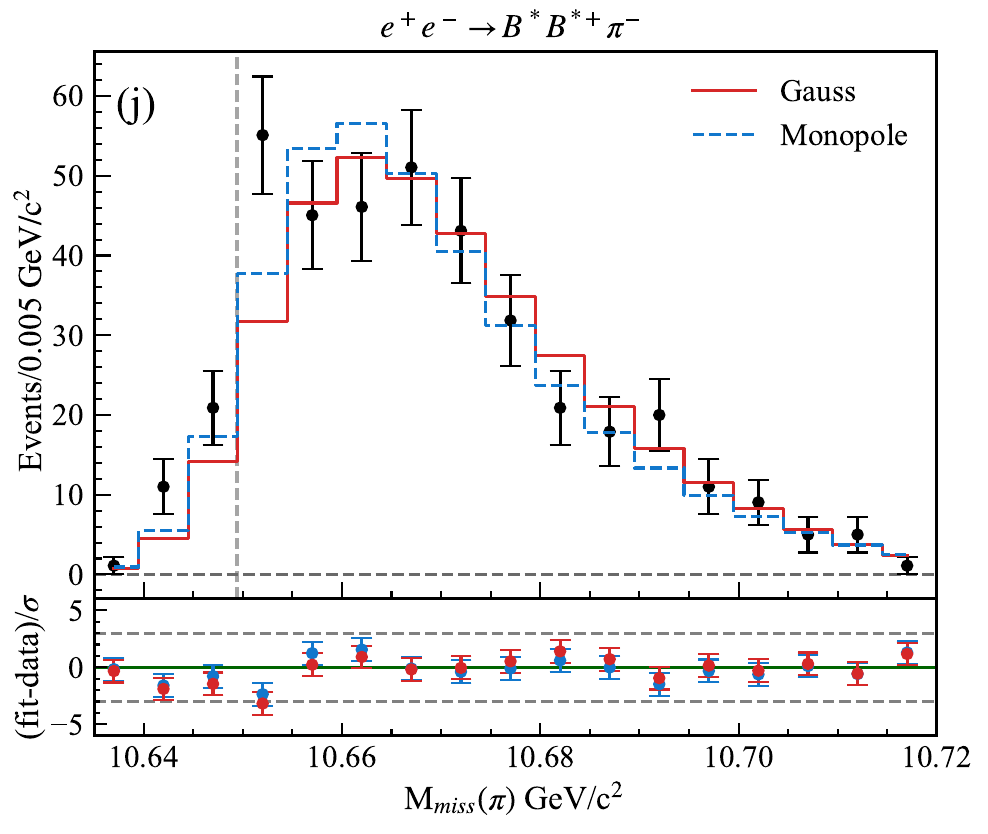}
    \includegraphics[width=0.28\linewidth]{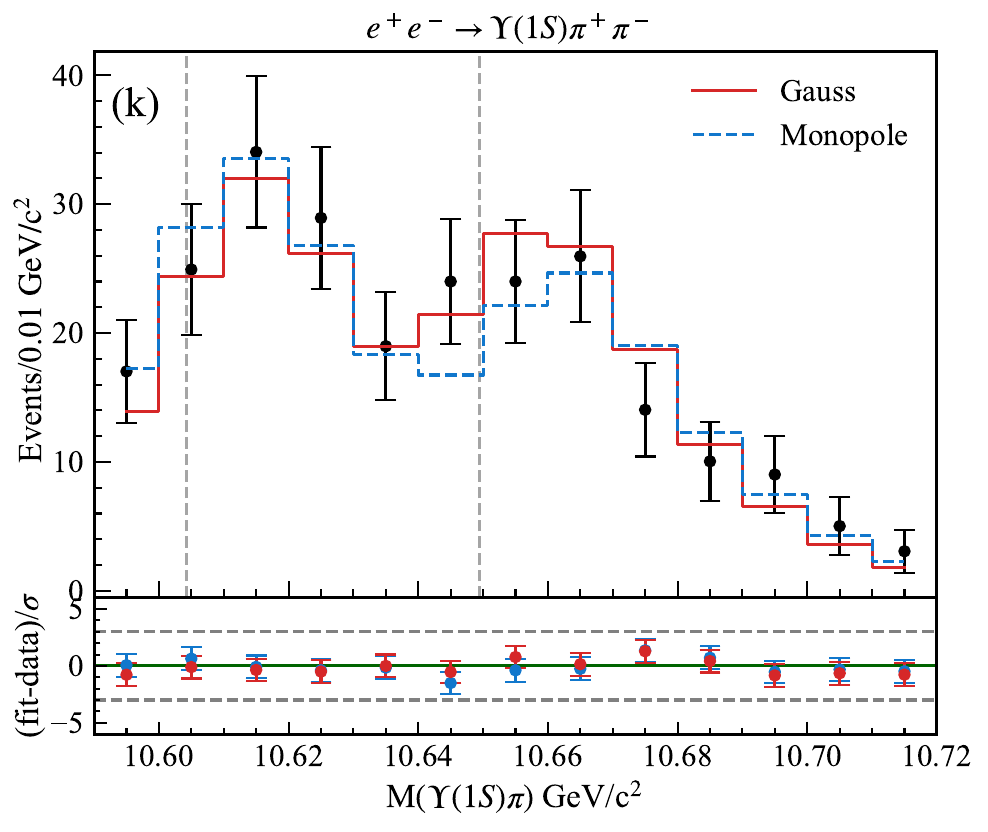}
    \includegraphics[width=0.28\linewidth]{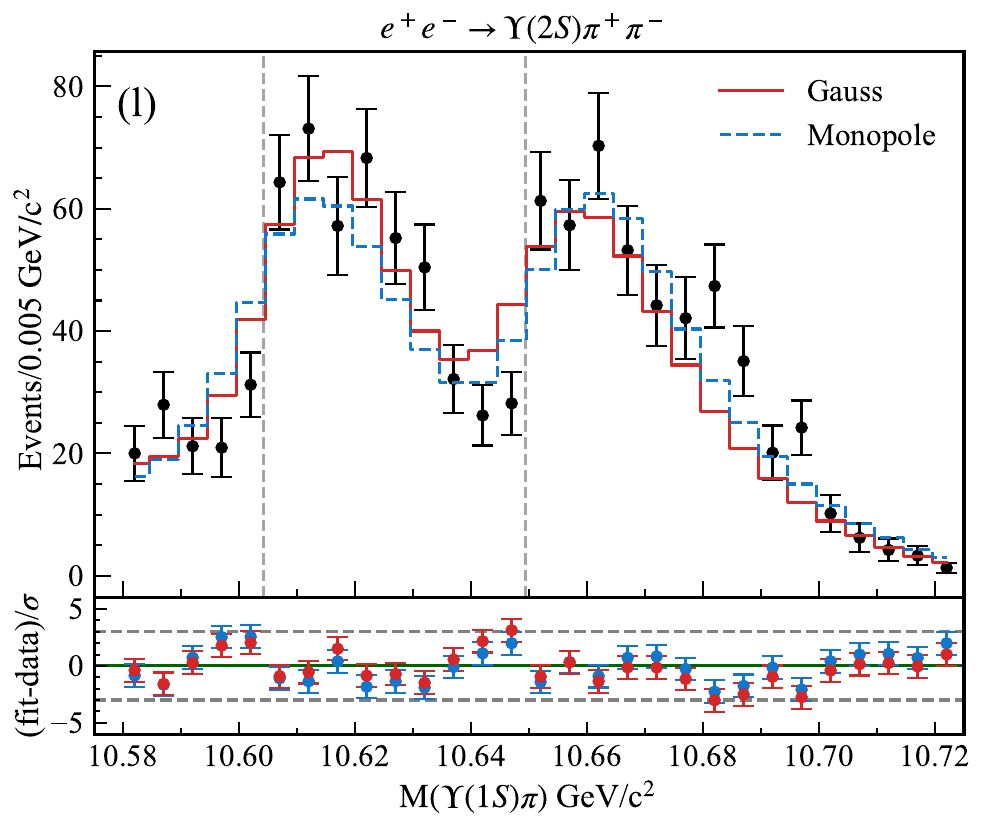}
    \includegraphics[width=0.28\linewidth]{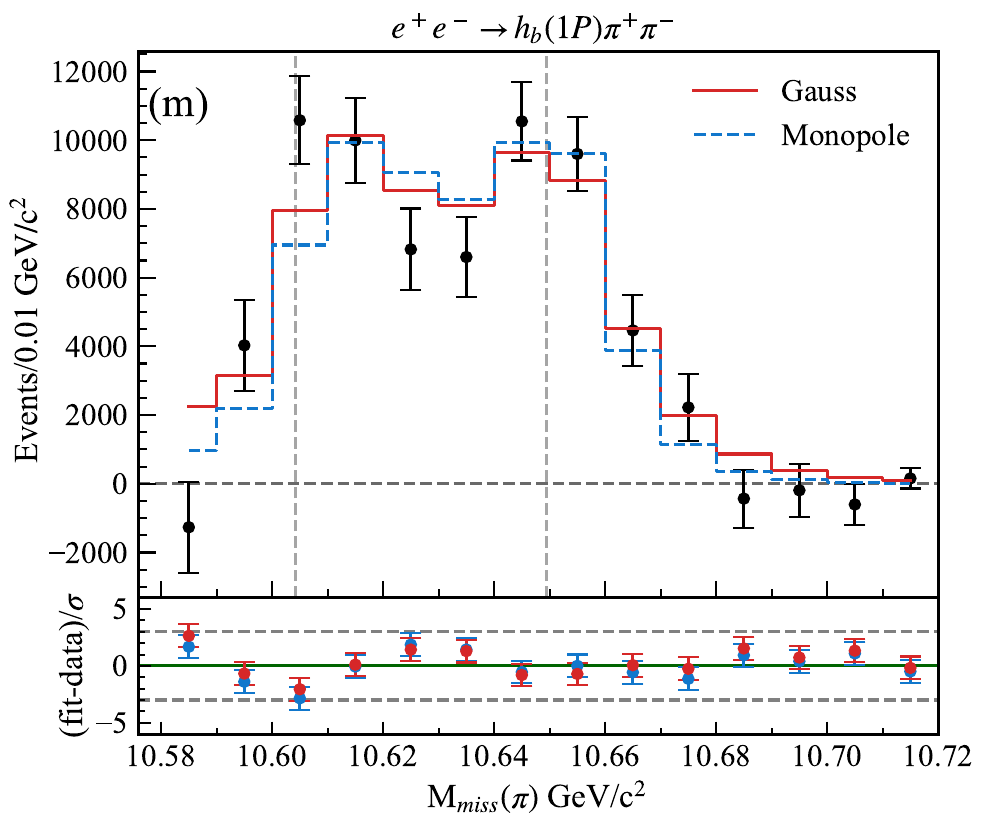}
    \includegraphics[width=0.28\linewidth]{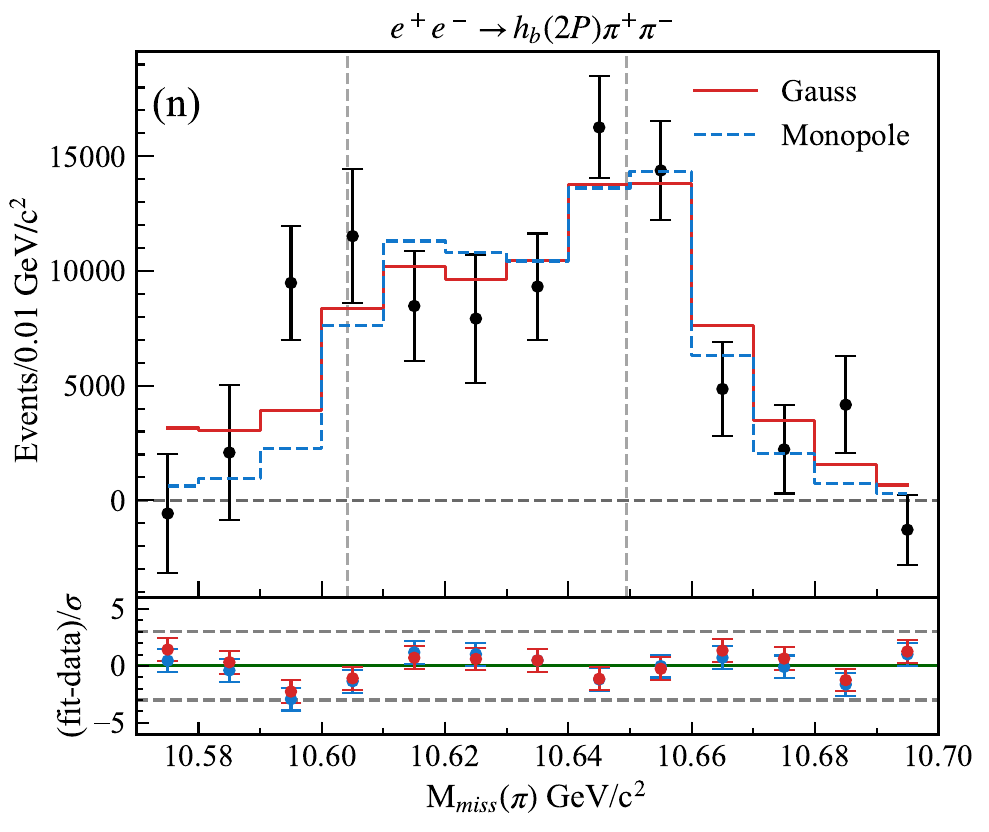}
    \caption{Line shapes of the $Z_c$ and $Z_b$ systems compared with experimental data after replacing the Gaussian form factor with a monopole form factor. Panels (a)–(h) show the invariant-mass distributions of the $Z_c$ system, while panels (i)–(n) display those of the $Z_b$ system. The red solid lines
    and blue dashed 
    lines represent the fitted results obtained with the Gaussian and monopole form factors, respectively.}
    \label{line_shape_Z2}
\end{figure*}

Replacing the two-point function in Eq.~(\ref{G_fun}) with the simplified expression in Eq.~(\ref{G_mono}), we refit the model. For simplicity, the analysis is performed within the HQSV scheme. The corresponding fit line shapes are displayed in Fig.~\ref{line_shape_Z2}. From the fitted parameters, we further extract the pole positions of the $Z_c$ and $Z_b$ systems, which are summarized in Tab.~\ref{z2_pole}. A comparison of the line shapes, the reduced chi-squares $\chi^2/\text{d.o.f}$, and the pole positions shows that the results are nearly unchanged when switching from the 
Gaussian form factor to the monopole one. This confirms that both choices describe the data equally well and demonstrates the robustness and stability of our model.
\begin{table}[hbt!]
\renewcommand{\arraystretch}{1.5}
    \centering
    \caption{Pole positions of the $Z_c$ and $Z_b$ systems with the monopole form factor.The units of mass and width are GeV.}
    \label{z2_pole}
    \begin{tabular}{p{2.5cm}<{\centering}p{2.5cm}<{\centering}p{2.5cm}<{\centering}}
    \hline\hline
   \textbf{R.S.}&  \textbf{HQSV}$(Z_c)$ & \textbf{HQSV}$(Z_b)$  \\
    \hline
   $(-,+)$  &  $3.875-0.012i$  & $10.597-0.025i$\\
   \hline
   $(-,-)$  &  $-$  & $10.645-0.012i$\\
   \hline
   \hline
   \end{tabular}  
\end{table}
